 \documentclass[
preprintnumbers,a4paper, amsmath, amssymb, floatfix,
nofootinbib, twocolumn, wrapfloat byrevtex]{revtex4}

 \usepackage{amsmath,mathrsfs,graphicx,epstopdf,placeins,float}
 \usepackage{booktabs}
 \usepackage{multirow}
 \usepackage{pstricks}

 \newcommand{\beq}[1]{\begin{equation}\label{#1}}
 \newcommand{\eeq}{\end{equation}}
 \newcommand{\bea}[1]{\begin{eqnarray}\label{#1}}
 \newcommand{\eea}{\end{eqnarray}}

\begin{document} 

 \title{Cosmological Complexity in K-essence}
 
 \author{Ai-chen Li $^{a,b}$}
\email{lac@emails.bjut.edu.cn, aichenli@icc.ub.edu}
\author{Xin-Fei Li$^{c}$}
\email{xfli@gxust.edu.cn}
\author{Ding-fang Zeng$^{d}$}
\email{dfzeng@bjut.edu.cn}
\author{Lei-Hua Liu}
\affiliation{\it ${}^a$ Institut de Ci\`encies del Cosmos, Universitat de Barcelona, Mart\'i i Franqu\`es 1, 08028 Barcelona, Spain}
\affiliation{\it ${}^b$ Departament de F\'isica Qu\`antica i Astrof\'isica, Facultat de F\'isica, Universitat de Barcelona, Mart\'i i Franqu\`es 1, 08028 Barcelona, Spain}
\affiliation{\it ${}^c$ School of Science, Guangxi University of Science and Technology, 545026 Liuzhou, China}
\affiliation{\it ${}^d$ Theoretical Physics Division, College of Applied Sciences, Beijing University of Technology, China}
\affiliation{\it ${}^e$ Department of Physics, College of Physics,
	Mechanical and Electrical Engineering,
	Jishou University, 416000 Jishou, China}

\begin{abstract}
We calculate the cosmological complexity under the framework of scalar curvature perturbations for a K-essence model with constant potential. In particular, the squeezed quantum states are defined by acting a two-mode squeezed operator which is characterized by squeezing parameters $r_k$ and $\phi_k$ on vacuum state. The evolution of these squeezing parameters are governed by the $Schr\ddot{o}dinger$ equation, in which the Hamiltonian operator is derived from the cosmological perturbative action. With aid of the solutions of $r_k$ and $\phi_k$, one can calculate the quantum circuit complexity between unsqueezed vacuum state and squeezed quantum states via the wave-function approach. One advantage of K-essence is that it allows us to explore the effects of varied sound speeds on evolution of cosmological complexity. Besides, this model also provides a way for us to distinguish the different cosmological phases by extracting some basic informations, like the scrambling time and Lyapunov exponent etc, from the evolution of cosmological complexity.

\end{abstract}
\pacs{}
\maketitle

\section{Introduction}

In recent years, the association between quantum information/computation theory and gravity has attracted a great amount of attentions. At the begining, this idea was motivated by the Anti-de Sitter/conformal field theory (AdS/CFT) \cite{Maldacena:1997re, Witten:1998qj, Gubser:1998bc}, especially the research about holographic entanglement entropy \cite{Ryu:2006bv}. Afterwards, another complementary physical quantity called quantum circuit complexity has been involved into the AdS/CFT dictionary. In particular, for a thermo-field double state  which is dual to an eternal asymptotic-AdS black hole \cite{Maldacena:2001kr}, \cite{Hartman:2013qma} indicates that the entanglement entropy fails to depict the growth behavior after reaching the thermal equilibrium for the Einstein-Rosen Bridge (ERB) behind the horizon. As an alternative way, \cite{Stanford:2014jda,Susskind:2014rva} propose that the growth behavior of ERB in black hole interior, particularly on late time, corresponds to the evolution of quantum circuit complexity between reference state and target state on AdS boundary. Specifically, this conjecture called complexity-volume (CV) supposes that the circuit complexity derived from the CFT on boundary is dual to the maximum volume of the ERB in bulk spacetime. Subsequently, \cite{Brown:2015lvg,Brown:2015bva} suggest another version of CV called complexity-action (CA) conjecture, which associates the circuit complexity on spacetime boundary to the gravitational action evaluated on a region of Wheeler-DeWitt  patch in the bulk spacetime. After that, many extensive studies on CV and CA conjectures have produced many profound results as shown in \cite{Carmi:2016wjl,Reynolds:2016rvl,Brown:2017jil,Alishahiha:2017hwg,Cano:2018aqi,Karar:2019bwy,Ling:2019ien,Hernandez:2020nem,Li:2020ark}.

Inspired by the holographic duality, more and more interests are focused on understanding the physics of quantum circuit complexity from the sides of quantum field theory \cite{Chapman:2017rqy,Jefferson:2017sdb,Hackl:2018ptj} or quantum mechanics \cite{Ali:2019zcj,Bhattacharyya:2020art}. As pointed out in \cite{Ali:2019zcj}, some elementary  information about a quantum chaotic system, like the scrambling time and Lyapunov exponent, can be captured by circuit complexity. The Fubini-Study approach is proposed in \cite{Chapman:2017rqy}, in order to measure the complexity is identified as the geodesic distance connecting the reference and the target states in group manifold (both the reference and the target states should be the coherent states of a specific group). Besides, for Gaussian quantum states, anther geometric way for calculating the quantum circuit complexity is given by Nielsen \cite{NielsenComplexity1,NielsenComplexity2,NielsenComplexity3}, and has been generalized to the context of QFT by \cite{Jefferson:2017sdb}, including the wave-function and the covariance matrix approachs respectively \cite{Ali:2018fcz}.

Recently, in field of cosmology, the application of quantum circuit complexity to scalar curvature perturbation on an expanding Friedmann-Lemaitre-Robertson-Walker (FLRW) background has been investigated by \cite{Bhattacharyya:2020rpy}. Similar to the definition of squeezed quantum states in inverted harmonic oscillator \cite{Barton:1984ey,TQO}, the two-mode squeezed state formalism and the corresponding differential equations in framework of cosmological perturbations are developed by \cite{Grishchuk:1990bj,Albrecht:1992kf,Martin:2019wta}. In \cite{Bhattacharyya:2020rpy}, the complexity of cosmological perturbations (hereafter we call this quantity as cosmological complexity) between unsqueezed vacuum state and squeezed quantum state is computated by using the wave-function approach. Their results uncover that during the inflation epoch the complexity is frozen inside the horizon, while it grows in an exponential way after the mode exits the horizon. And then the universe de-complexifies during the subsequent radiation epoch and eventually the complexity is frozen after horizon re-entry. Since then,  \cite{Bhattacharyya:2020kgu} explores the cosmological complexity for both expanding and contracting FLRW backgrounds with varied equation of state $w$. Besides, the cosmological complexity is also studied in some typical cosmological models which are alternative theories to the cosmic inflation scenario, like ekpyrosis and bouncing cosmology \cite{Lehners:2020pem,Bhargava:2020fhl}.

In this paper, our aim is to investigate the cosmological complexity in K-essence cosmology models \cite{ArmendarizPicon:1999rj,Garriga:1999vw,Jorge:2007zz}, which are typically described by a large class of higher-order (non-quadratic) scalar kinetic terms. K-essence as an important model to drive cosmic inflation, manifesting the many aspects of advantages that inflationary evolution is driven by higher-order scalar kinetic terms only and inflation starts from very generic initial conditions without the help of potential terms. Moreover, the dynamical attractor solutions derived from the K-essence can avoid fine-tuning of parameters and anthropic arguments in explaining the accelerated expansion of the universe at present \cite{ArmendarizPicon:2000ah,Malquarti:2003hn,Malquarti:2003nn}. Our motivations come from the following aspects: Firstly, in \cite{Bhattacharyya:2020rpy} the cosmological complexity is considered in scalar curvature perturbations with constant sound speed, i.e. $c^2_S=1$. As a step forward, it is valuable to consider the effects of varied sound speed on the evolution of cosmological complexity. This purpose could be achieved in perturbative theory of K-essence\cite{Garriga:1999vw}, the varied $c^2_S$ is included in scalar curvature perturbations since the higher order corrections on the canonical momentum for scalar fields. Secondly, for the K-essence model given by \cite{Jorge:2007zz}, the enriched cosmological phases could be observed in same physical parameters with different initial conditions. Although the differences of these cosmological phases are reflected by the equations of state and scale factor, we also expect that the evolution of the cosmological complexity could provide some information to distinguish these cosmological phases.

Our work is structured as follows. In section \ref{Kessence}, an specific K-essence cosmology model and the corresponding perturbative theories are reviewed. In section \ref{squeezedSol}, by combining the definition of squeezed quantum states with the perturbative actions given in section \ref{Kessence}, we derive the differential equations governing the evolution of squeezing parameter $r_k$ and squeezing angle $\phi_k$, and the corresponding numerical solutions are obtained. The complexity between unsqueezed vacuum state and squeezed quantum states are computed through the wave-function approach in section \ref{SqueeComplex}. We give conclusions and some future directions in the last section.

\section{K-essence models and the corresponding cosmological perturbations \label{Kessence}}

\subsection{K-essence cosmology}

The K-essence cosmology are described by coupling a scalar field to Einstein gravity \cite{ArmendarizPicon:1999rj, Garriga:1999vw}
\begin{align}
\label{KessenceAct}
&S=\frac{1}{2}\int d^{4}x\sqrt{-g}R+\int d^{4}x\sqrt{-g}P(X,\varphi)
\end{align}
in which the Lagrangian $P(X,\varphi)$ is allowed to have a dependence on higher-order powers of the kinetic term $X=-\frac{1}{2} g^{\mu\nu}\partial_\mu \varphi \partial_\nu \varphi$. In $\eqref{KessenceAct}$, why we denote the Lagrangian for the scalar field as $P(X,\varphi)$ because it plays the role of pressure, as shown in $\eqref{EMToFluid}$. Note that we take the convention $8\pi G=c=1$, for the convenience of calculations. 

The Einstein field equations could be obtained by varying the Lagrangian $\eqref{KessenceAct}$ with respect to metric tensor,
\begin{align}
\label{KessEinFie}
&R_{\mu\nu}-\frac{1}{2}g_{\mu\nu}R=T_{\mu\nu}\\
\nonumber
&T_{\mu\nu}=\frac{\partial P(X,\varphi)}{\partial X}\partial_{\mu}\varphi\partial_{\nu}\varphi+g_{\mu\nu}P(X,\varphi)
\end{align} 
Decomposing the above energy momentum tensor into the form of a perfect fluid, we find
\begin{align}
\label{EMToFluid}
&T_{\mu\nu}=\mathcal{E}u_{\mu}u_{\nu}+P\big(g_{\mu\nu}+u_{\mu}u_{\nu}\big)
\end{align}
in which the 4-velocity is
\begin{align}
&u_{\mu}=\frac{\partial_{\mu}\varphi}{(2X)^{1/2}}
\end{align}
and the energy density is given by
\begin{align}
\label{EnerDensity}
&\mathcal{E}=2X\cdot\frac{\partial P}{\partial X}-P
\end{align}
In a background of flat Friedman-Lemaitre-Robertson-Walker (FLRW) universe,
\begin{align}
\label{FLRWmetric}
&ds^{2}=-dt^{2}+a(t)^{2}\delta_{ij}dx^{i}dx^{j}
\end{align}
the following independent equations are found by plugging $\eqref{FLRWmetric}$ into the Einstein field equation $\eqref{KessEinFie}$,
\begin{align}
\label{Friedmann1}
&~~3H^{2}=\mathcal{E}\\
\label{Friedmann2}
&-2\dot{H}=\mathcal{E}+P
\end{align}
in which $H=\frac{\dot{a}}{a}$ represents the Hubble constant. And then, by varying the Lagrangian with respect to $\varphi$, the equation of motion for scalar field reads
\begin{align}
\label{CovaScalar}
&\nabla_{\nu}\big(\frac{\partial P(X,\varphi)}{\partial X}g^{\mu\nu}\partial_{\mu}\varphi\big)+\frac{\partial P(X,\varphi)}{\partial\varphi}=0
\end{align}
expand $\eqref{CovaScalar}$ explicitly, we give
\begin{align}
\nonumber
&\ddot{\varphi}\frac{\partial P(X,\varphi)}{\partial X}+3H\dot{\varphi}\frac{\partial P(X,\varphi)}{\partial X}\\
\label{ExpScalar}
&\quad\quad\quad\quad\quad+\dot{\varphi}\frac{d}{dt}\big(\frac{\partial P(X,\varphi)}{\partial X}\big)=\frac{\partial P(X,\varphi)}{\partial\varphi}
\end{align}
Note that the $\eqref{ExpScalar}$ could also be derived from the continuity equation $\dot{\mathcal{E}}+3H(\mathcal{E}+P)=0$. It should be indicated that there are only two independent equations among $\eqref{Friedmann1},\eqref{Friedmann2},\eqref{CovaScalar}$.

\begin{figure}[ht]
	\begin{center}
		\includegraphics[scale=0.34]{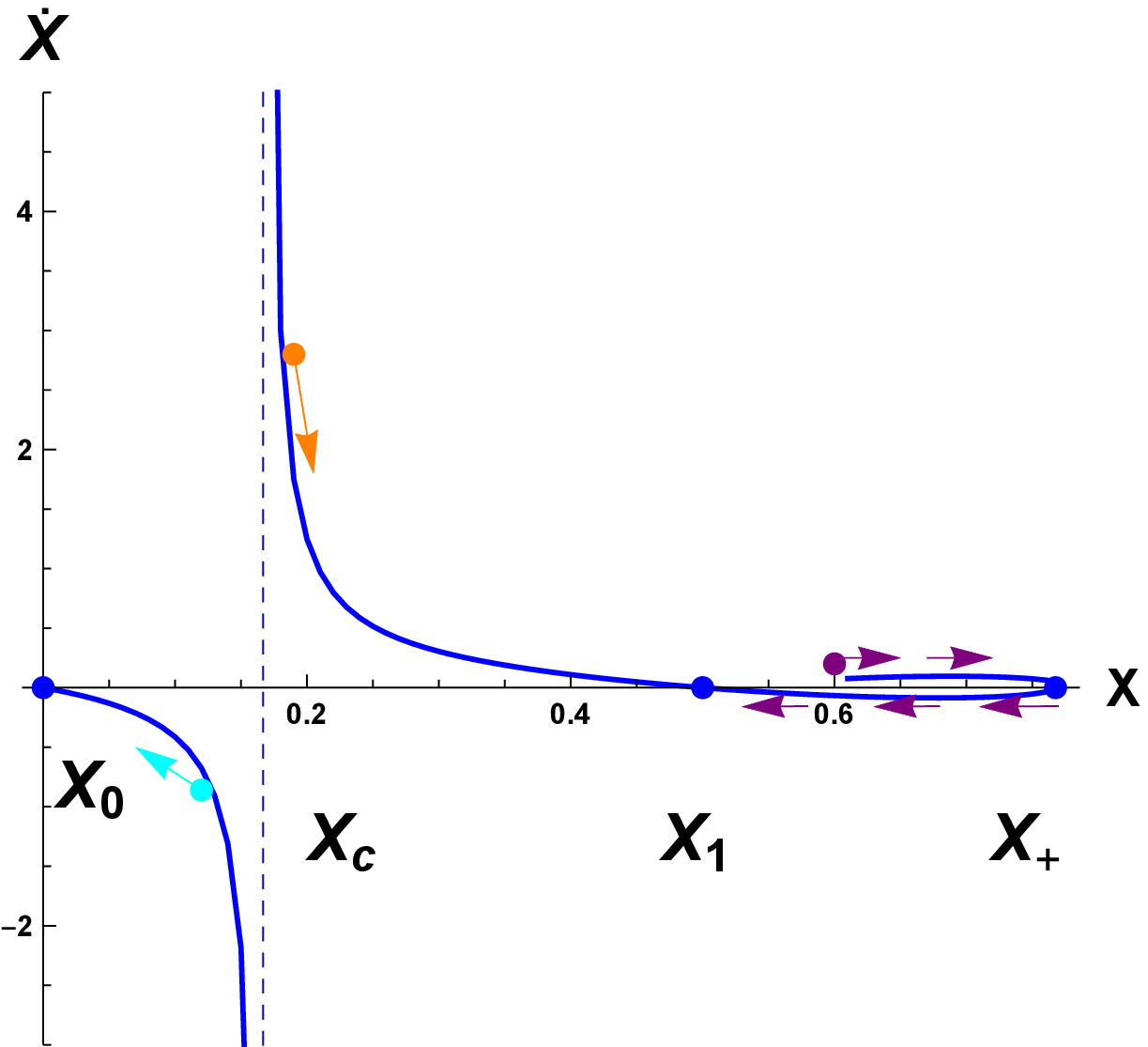}
		\includegraphics[scale=0.34]{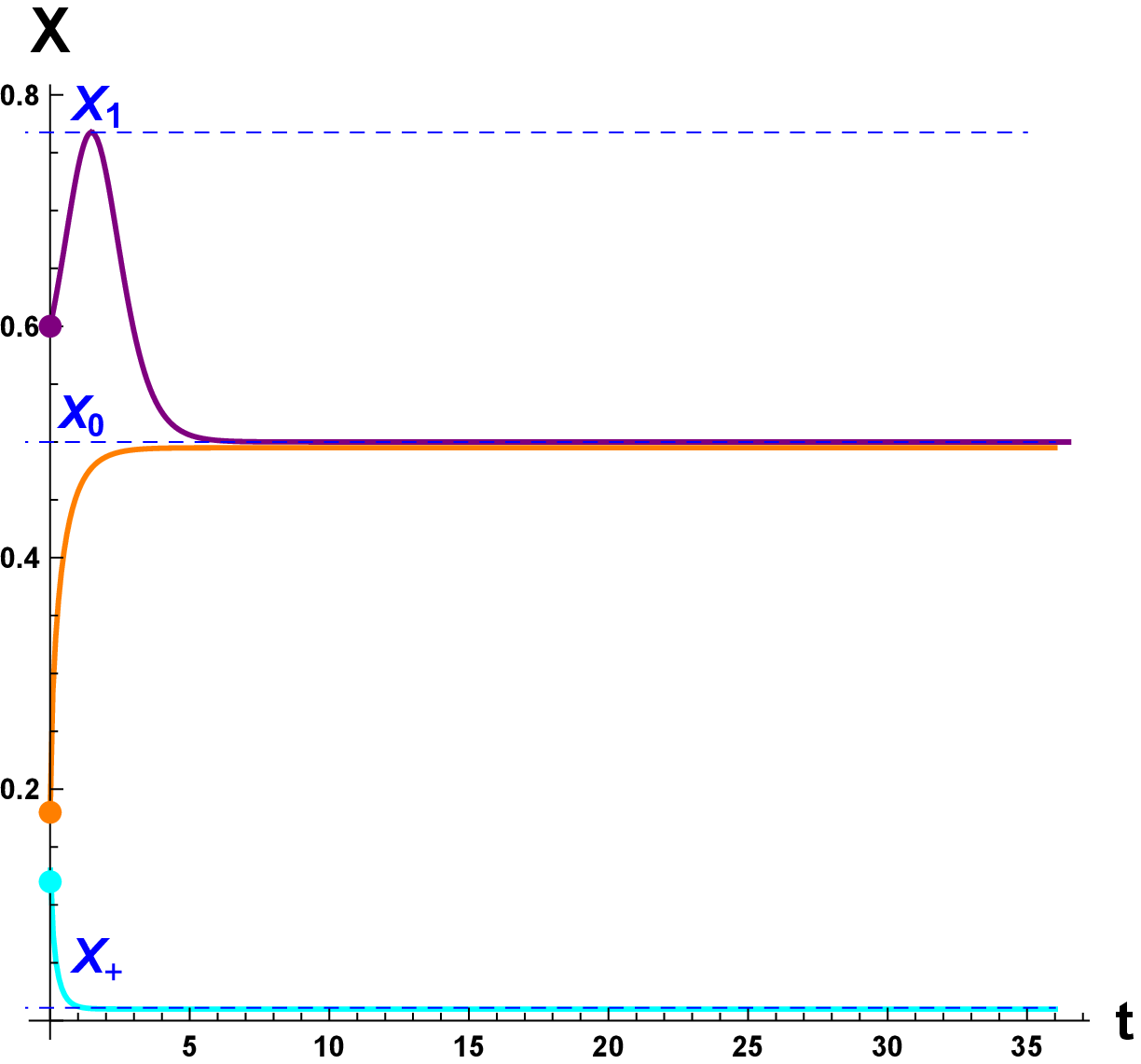}\\
		\includegraphics[scale=0.34]{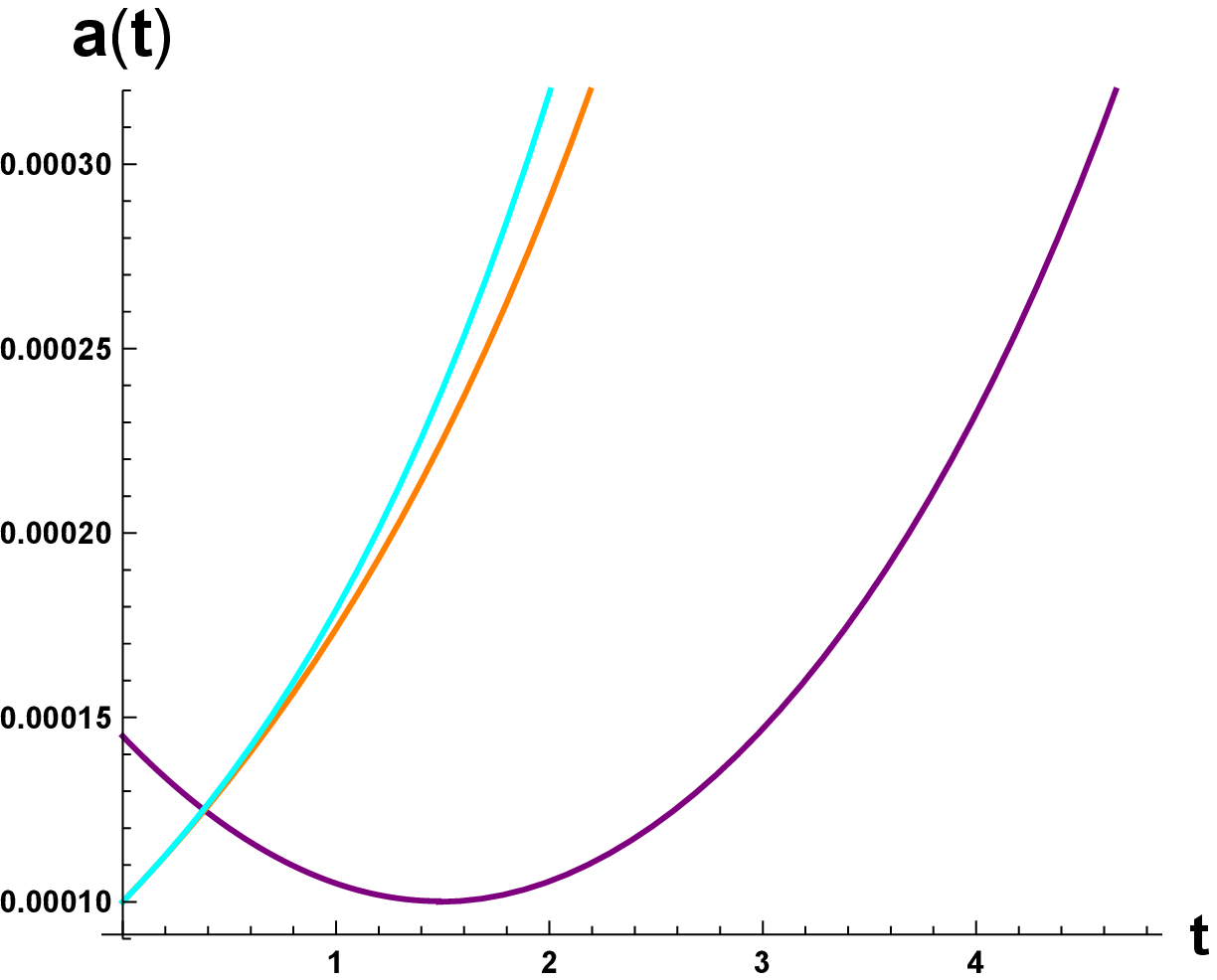}
		\includegraphics[scale=0.34]{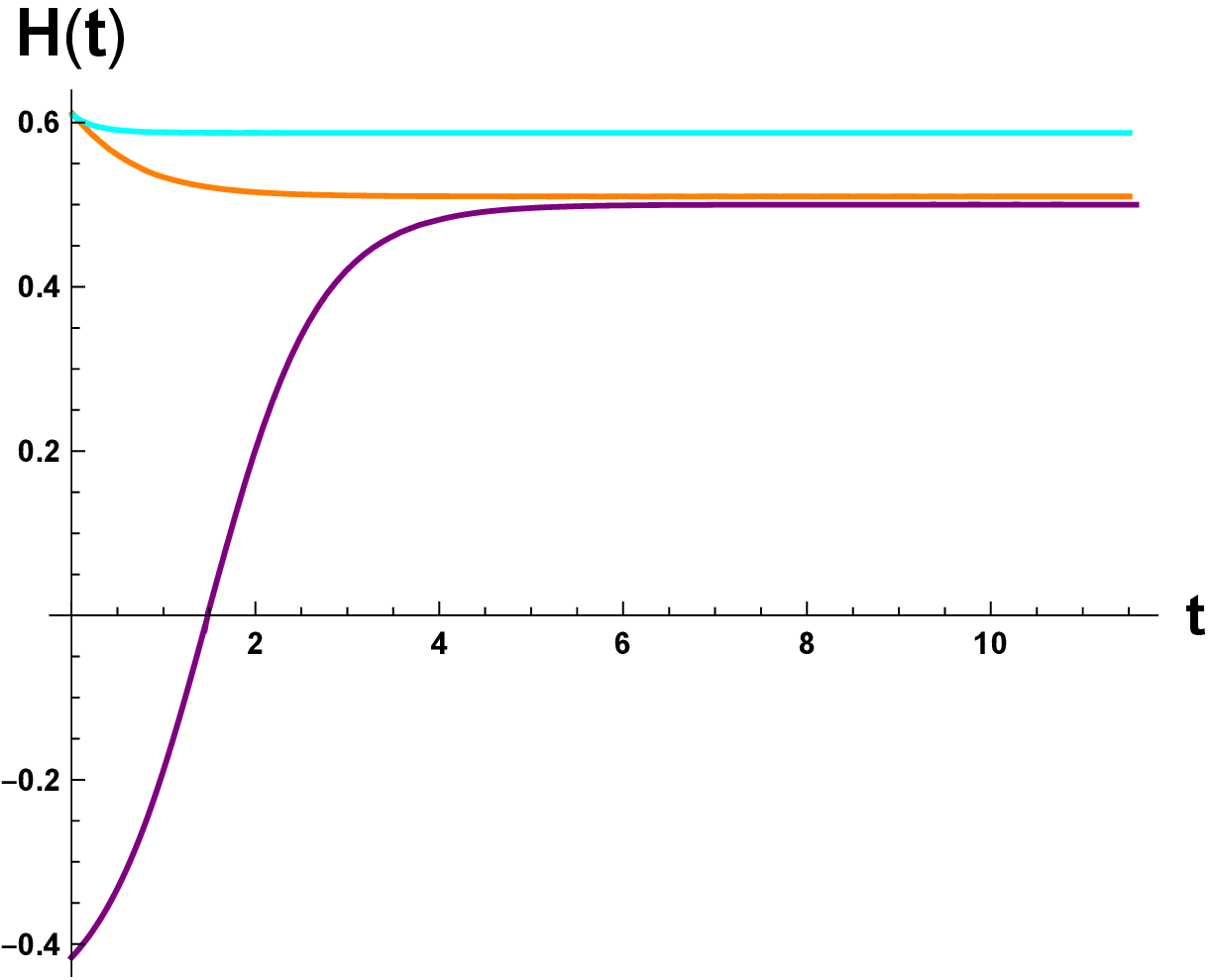}
		\caption{(color online). According to the expression of $\eqref{NewSpecEOMsVarphi}$, we plot the variation of $\dot{X}$ with respect to $X$ in case of the representative parameters $V_0=1~,C_0=-2$. While, the change of $X, a, H$ with time are shown by solving $\eqref{SpecFried1}, \eqref{SpecEOMsVarphi}$ numerically. The corresponding initial conditions in domains $[X_0, X_c)$, $(X_c, X_1)$, $(X_1, X_+)$ are denoted with cyan, orange, purple points.}
		\label{EOMsXRHS}
	\end{center}
\end{figure}

\begin{figure}[ht]
	\begin{center}
		\includegraphics[scale=0.34]{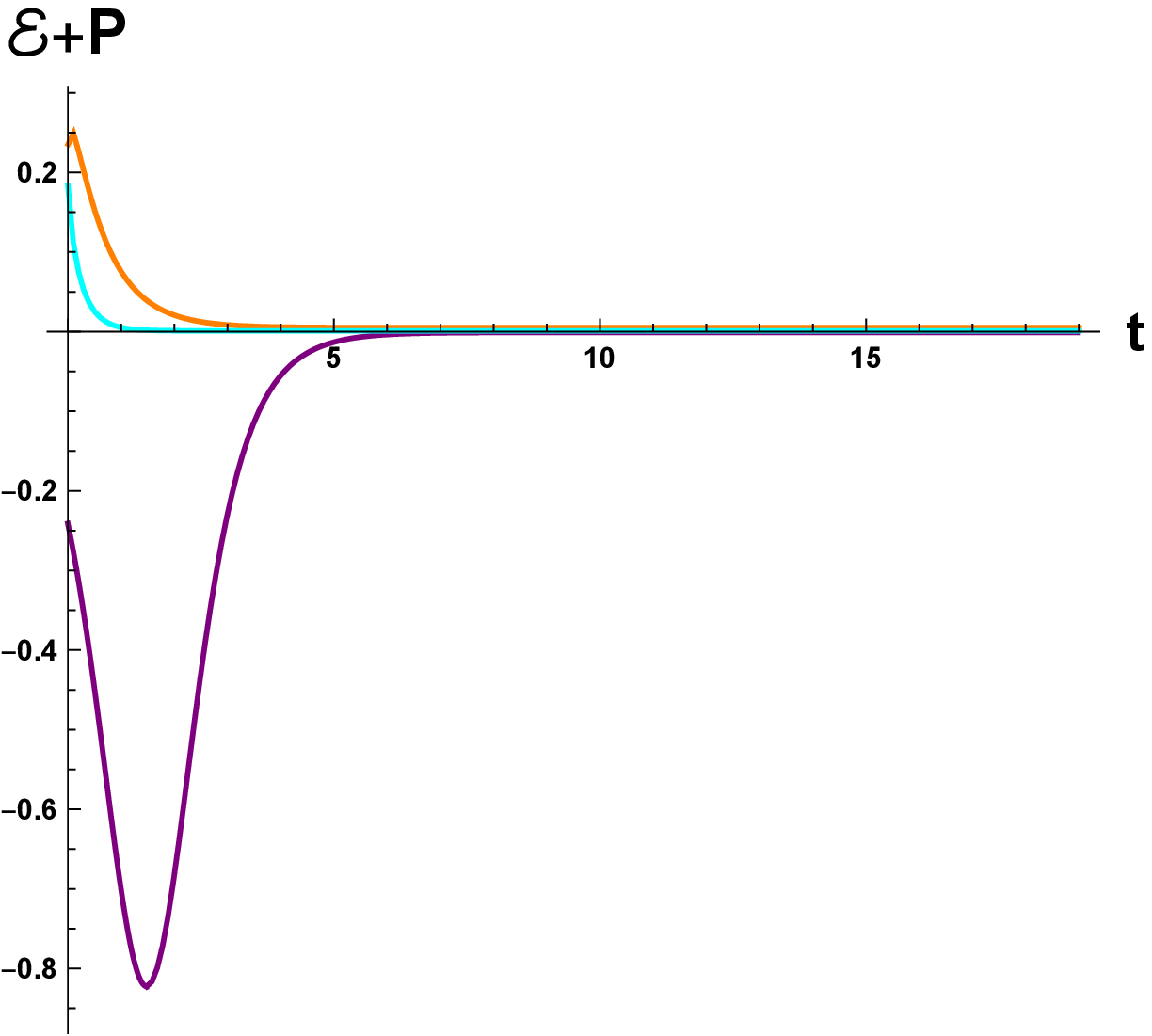}
		\includegraphics[scale=0.34]{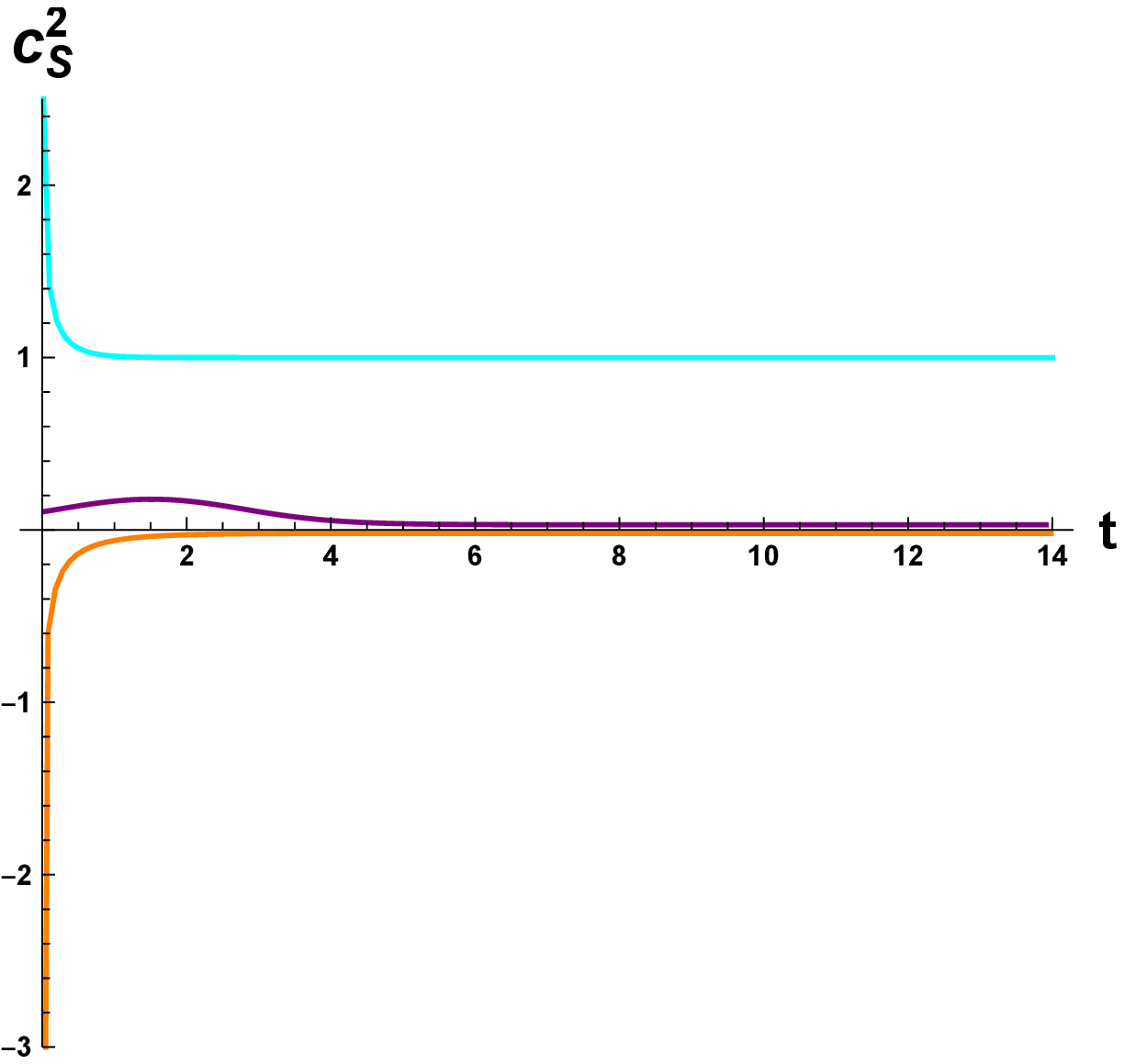}
		\caption{(color online). The variation of $\mathcal{E}+P$, $c^2_S$ as the time increases, which corresponds to the numerical results given in Fig.\ref{EOMsXRHS}. }
		\label{EnerMoCSsqureSol}
	\end{center}
\end{figure}

In this paper, we will consider the quadratic purely kinetic Lagrangian with constant potential $V(\varphi)$ \cite{Jorge:2007zz}, namely
\begin{align}
\label{QuadraKineWithCon}
&P(X,\varphi)=X+\frac{C_{0}}{2}X^{2}-V_{0}
\end{align}
Plug $\eqref{QuadraKineWithCon}$ into $\eqref{Friedmann1}$ and $\eqref{ExpScalar}$, we get the differential equations which controll the evolution of scale factor $a(t)$ and classical scalar field $\varphi(t)$
\begin{align}
\label{SpecFried1}
&3H^{2}=X+\frac{3}{2}C_{0}X^{2}+V_{0}\\
\label{SpecEOMsVarphi}
&\dot{X}=-6HX\frac{1+C_{0}X}{1+3C_{0}X}
\end{align}
From $\eqref{EnerDensity}$, $\eqref{QuadraKineWithCon}$, the effective speed of sound is expressed as
\begin{align}
&c^2_S=\frac{dP/dt}{d\mathcal{E}/dt}=\frac{\partial P/\partial X}{\partial\mathcal{E}/\partial X}=\frac{1+C_{0}X}{1+3C_{0}X}
\end{align}
In $\eqref{SpecEOMsVarphi}$, let us eliminate $H$ by using $\eqref{SpecFried1}$
\begin{align}
\label{NewSpecEOMsVarphi}
&\dot{X}=\pm 6X\sqrt{\frac{\big(X+\frac{3}{2}C_{0}X^{2}+V_{0}\big)}{3}}\frac{1+C_{0}X}{1+3C_{0}X}
\end{align}
Note that the "+" and "-" branches in $\eqref{NewSpecEOMsVarphi}$ correspond to the $H<0$ and $H>0$ respectively. Here we will restrict our attention on the case of $V_0>0~,C_0<0$ in order to give the enriched cosmological phases in different initial conditions. The sqrt in $\eqref{NewSpecEOMsVarphi}$ and condition $X=\frac{1}{2}(\dot{\varphi})^{2}>0$ imply that the $X$ is well-defined only in regions $[0,~\frac{\sqrt{1-6C_0 V_0}-1}{3C_0}]$. As shown by the left panel of Fig.\ref{EOMsXRHS}, the right-hand side of $\eqref{NewSpecEOMsVarphi}$ vanishes at three places $X_0=0$, $X_1=-\frac{1}{C_0}$ and $X_+=\frac{\sqrt{1-6C_0 V_0}-1}{3C_0}$. Meanwhile, this dynamical system is not defined at $X_c=-\frac{1}{3C_0}$, which is a terminating singularity but not a curvature singularity.
After solving this dynamical system $\eqref{SpecFried1}$-$\eqref{NewSpecEOMsVarphi}$ numerically, as shown by Fig.\ref{EOMsXRHS}, the various cosmological phases are shown when $X(t)$ is in different initial positions. In the domain $[X_0,X_c)$, as plotted by the cyan curves, the usual inflationary de Sitter phase is observed as $X(t)$ approaches the $X_0$. From Fig.\ref{EnerMoCSsqureSol}, it is easy to see that this phase has a well-defined behavior, namely $\mathcal{E}+P>0$ and $c^2_S>0$. However, although the orange curves whose initial value of $X(t)$ is in domain $(X_c, X_1)$ has the de Sitter behaviour as well, it exhibits a negative velocity of sound , $c^2_S<0$, which implies an exotic state. Finally, as displayed by the purple curves, the $X(t)$ moves toward the $X_+$ from an initial position in the region $(X_1, X_+)$, and then moves in opposite direction after reaching this point. This motion mode corresponds to a bounce phase in cosmology which transits from a phantom state to another phantom state.
It is necessary to point out that the evolution of $X(t)$ is controlled by the "+" and "-" branches respectively, in process of approaching and leaving the bouncing point $X_+$ respectively. 

\subsection{Perturbative actions}

Under the inhomogeneous quantum fluctuations, the scalar $\varphi$ has the following ansatz
\begin{align}
\label{PerVar}
&\varphi(t,\vec{x})=\bar{\varphi}(t)+\delta\varphi(t,\vec{x})
\end{align}
Meanwhile, in the longitudinal gauge \cite{Mukhanov:1990me}, the perturbed spacetime metric is written as
\begin{align}
\label{PerMetric}
&ds^{2}=-\big(1+2\Phi(t,\vec{x})\big)dt^{2}+a^{2}(t)\big(1-2\Phi(t,\vec{x})\big)\delta_{ij}dx^{i}dx^{j}
\end{align}
After substituting these perturbative ansatzs into $\eqref{KessenceAct}$, the perturbative action in second order is \cite{Garriga:1999vw}
\begin{align}
\label{PerturActionv1}
&\hspace{-7mm}S=\int dtd^{3}x\big\{\xi\hat{\Box}\dot{\zeta}-\frac{H^{2}c_{S}^{2}}{2a^{3}(\mathcal{E}+P)}\xi{\hat{\Box}}^2\xi+\frac{a(\mathcal{E}+P)}{2H^{2}}\zeta\hat{\Box}\zeta\big\}
\end{align}
where the new variables $\xi$ and $\zeta$ are associated to the $\delta\varphi$ and $\Phi$ through the definitions
\begin{align}
\label{definexi}
&\Phi a=\frac{1}{2}H\xi\\
\label{definezeta}
&\frac{\delta\varphi}{\dot{\varphi}}=\frac{\zeta}{H}-\frac{1}{2a\xi}
\end{align}
Besides, the perturbative Einstein equations in linearized order imply the two independent equations
\begin{align}
\label{PerEin00}
&\frac{1}{a^{2}}\hat{\Box}\Phi-3H\dot{\Phi}-3H^{2}\Phi=\frac{1}{2}\delta T_{0}^{0}\\
\label{PerEin0i}
&\partial_{i}\big(\dot{\Phi}+H\Phi\big)=\frac{1}{2}\delta T_{i}^{0}\\
\nonumber
&\delta T_{0}^{0}=\frac{\partial\mathcal{E}}{\partial X}\delta X+\frac{\partial\mathcal{E}}{\partial\varphi}\delta\varphi\\
\nonumber
&\quad~~=\frac{\mathcal{E}+P}{c_{S}^{2}}\big(\frac{\delta\varphi}{\dot{\varphi}}-\Phi\big)-3H(\mathcal{E}+P)\frac{\delta\varphi}{\dot{\varphi}}\\
\nonumber
&\delta T_{i}^{0}=(\mathcal{E}+P)\partial_{i}\bigg(\frac{\delta\varphi}{\dot{\varphi}}\bigg)
\end{align}
Substituting $\eqref{definexi}$, $\eqref{definezeta}$ into $\eqref{PerEin00}$, $\eqref{PerEin0i}$ and combining with the classical EOMs $\eqref{Friedmann1}-\eqref{ExpScalar}$, we get the following equations for $\xi$ and $\zeta$
\begin{align}
\label{eqsdotxi}
&\dot{\xi}=\frac{a(\mathcal{E}+P)}{H^{2}}\zeta\\
\label{eqsdotzeta}
&\dot{\zeta}=\frac{c_{S}^{2}H^{2}}{a^{3}(\mathcal{E}+P)}\hat{\Box}\xi
\end{align}
The action $\eqref{PerturActionv1}$ could be further simplified via the equation $\eqref{eqsdotzeta}$,
\begin{align}
\label{PerturActionv2}
&S=\frac{1}{2}\int d\eta d^{3}x~z^{2}\big(\zeta^{\prime2}+c_{S}^{2}\zeta\hat{\Box}\zeta\big)
\end{align}
in which we have used the conformal time $\eta=\int \frac{dt}{a(t)}$ and prime denotes the derivative with respect to $\eta$. Meanwhile, the variable $z$ is
\begin{align}
\label{Zvariable}
&z=\frac{a(\mathcal{E}+P)^{1/2}}{c_{S}H}
\end{align}
By introducing the Mukhanov variable $v=z\zeta$, one can rewrite the action $\eqref{PerturActionv2}$ as
\begin{align}
\label{PerLagranKessen}
&S=\frac{1}{2}\int d\eta d^{3}x\bigg(v^{\prime2}+c_{S}^{2}v\hat{\Box}v+\frac{z^{\prime\prime}}{z}v^{2}\bigg)
\end{align}

\section{The squeezed quantum states for cosmological perturbations \label{squeezedSol}}

By using the integration by parts, the action $\eqref{PerLagranKessen}$ are transformed into
\begin{align}
\nonumber
&S=\int d\eta L=\frac{1}{2}\int d\eta d^{3}x\bigg(v^{\prime2}-c_{S}^{2}(\partial_{i}v)^{2}\\
\label{PerLagKessenCano}
&\quad\quad\quad\quad\quad\quad+\big(\frac{z^{\prime}}{z}\big)^{2}v^{2}-2\frac{z^{\prime}}{z}v^{\prime}v\bigg)
\end{align}
In this paper, we shall restrict our attention to the case of flat universe, so we have set $K=0$ in $\eqref{PerLagKessenCano}$. From $\eqref{PerLagKessenCano}$, the canonical momentum is defined as
\begin{align}
&\pi(\eta,\vec{x})=\frac{\delta L}{\delta v^{\prime}(\eta,\vec{x})}=v^{\prime}-\frac{z^{\prime}}{z}v
\end{align}
So the Hamiltonian $H=\int d^{3}x(\pi v^{\prime}-\mathcal{L})$ is constructed as
\begin{align}
\label{HamilKessenCano}
&H=\frac{1}{2}\int d^{3}x\big[\pi^{2}+c_{S}^{2}(\partial_{i}v)^{2}+\frac{z^{\prime}}{z}(v\pi+\pi v)\big)\big]
\end{align}
By means of the method of second quantization, the field $v(\eta,\vec{x}),~\pi(\eta,\vec{x})$ are promoted to operators $\hat{v}(\eta,\vec{x}),~\hat{\pi}(\eta,\vec{x})$. Similar to the quantum mechanics of inverted harmonic oscillator \cite{Barton:1984ey}, we suppose the $\hat{v}(\eta,\vec{x})$ and $\hat{\pi}(\eta,\vec{x})$ have the following decomposition in Fourier space
\begin{align}
&\hspace{-2mm}\hat{v}(\eta,\vec{x})=\int\frac{d^{3}k}{(2\pi)^{3/2}}\frac{1}{\sqrt{2k}}\big(\hat{c}_{-\vec{k}}^{\dagger}v_{k}^{\star}(\eta)+\hat{c}_{\vec{k}}v_{k}(\eta)\big)e^{i\vec{k}\cdot\vec{x}}\\
&\hspace{-2mm}\hat{\pi}(\eta,\vec{x})=i\int\frac{d^{3}k}{(2\pi)^{3/2}}\sqrt{\frac{k}{2}}\big(\hat{c}_{-\vec{k}}^{\dagger}u_{k}^{\star}(\eta)-\hat{c}_{\vec{k}}u_{k}(\eta)\big)e^{i\vec{k}\cdot\vec{x}}
\end{align}
in which $\hat{c}_{-\vec{k}}^{\dagger}$ and $\hat{c}_{\vec{k}}$ are the creation and annihilation operators respectively. Through choosing an appropriate normalization condition for mode functions $u_k(\eta),~v_k(\eta)$, besides the necessary uncertainty relation
\begin{align}
&[v(\eta,\vec{x}),v^{\prime}(\eta,\vec{y})]\big\vert_{\eta=\eta_{0}}=i\delta^{3}(\vec{x}-\vec{y})
\end{align}
the Hamiltonian operator could be also simplified as
\begin{align}
\nonumber
&\hat{H}=\int d^{3}k\hat{\mathcal{H}}_{k}=\int d^{3}k\big\{\frac{k}{2}(c_{S}^{2}+1)\hat{c}_{-\vec{k}}^{\dagger}\hat{c}_{-\vec{k}}\\
\nonumber
&\quad~~+\frac{k}{2}(c_{S}^{2}+1)\hat{c}_{\vec{k}}\hat{c}_{\vec{k}}^{\dagger}+\big(\frac{k}{2}(c_{S}^{2}-1)+\frac{iz^{\prime}}{z}\big)\hat{c}_{\vec{k}}^{\dagger}\hat{c}_{-\vec{k}}^{\dagger}\\
\label{HamWithCs}
&\quad~~+\big(\frac{k}{2}(c_{S}^{2}-1)-\frac{iz^{\prime}}{z}\big)\hat{c}_{\vec{k}}\hat{c}_{-\vec{k}}\big\}
\end{align}
Just as the inverted harmonic oscillator, given the quadratic Hamiltonian, the unitary evolution operator could be factorized in the following form \cite{Grishchuk:1990bj, Albrecht:1992kf}
\begin{align}
\label{UniOperator}
&\hat{\mathcal{U}}_{\vec{k}}(\eta,\eta_{0})=\hat{\mathcal{S}}_{\vec{k}}(r_{k},\phi_{k})\hat{\mathcal{R}}_{\vec{k}}(\theta_{k})
\end{align}
In $\eqref{UniOperator}$, the $\hat{\mathcal{R}}_{\vec{k}}$ is the two-mode rotation operator with the following definition
\begin{align}
&\hat{\mathcal{R}}_{\vec{k}}(\theta_{k})=\exp\big[-i\theta_{k}(\eta)\big(\hat{c}_{\vec{k}}\hat{c}_{\vec{k}}^{\dagger}+\hat{c}_{-\vec{k}}^{\dagger}\hat{c}_{-\vec{k}}\big)\big]
\end{align}
where the $\theta_k (\eta)$ is the rotation angle. Meanwhile, $\hat{\mathcal{S}}$ is the two-mode squeeze operator defined as
\begin{align}
\label{SqueeOpeVo}
&\hspace{-5mm}\hat{S}_{\vec{k}}(r_{k},\phi_{k})=\exp\big[r_{k}(\eta)\big(e^{-2i\phi_{k}(\eta)}\hat{c}_{\vec{k}}\hat{c}_{-\vec{k}}-e^{2i\phi_{k}(\eta)}\hat{c}_{-\vec{k}}^{\dagger}\hat{c}_{\vec{k}}^{\dagger}\big)\big]
\end{align}
where $r_k(\eta)$ and $\phi_k(\eta)$ represent the squeezing parameter and squeezing angle respectively. Note that we will ignore the effects of rotation operators, because it only produces an irrelevant phase when acting on the initial vacuum state.

\begin{figure}[ht]
	\begin{center}
		\includegraphics[scale=0.33]{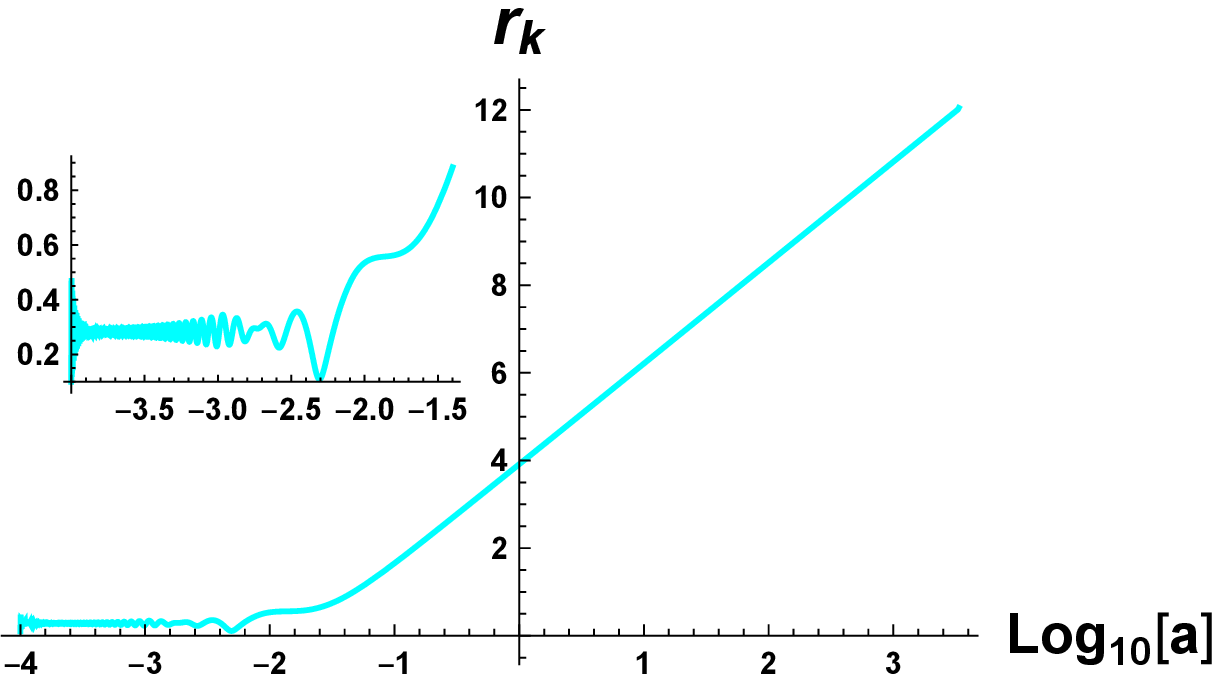}
		\includegraphics[scale=0.33]{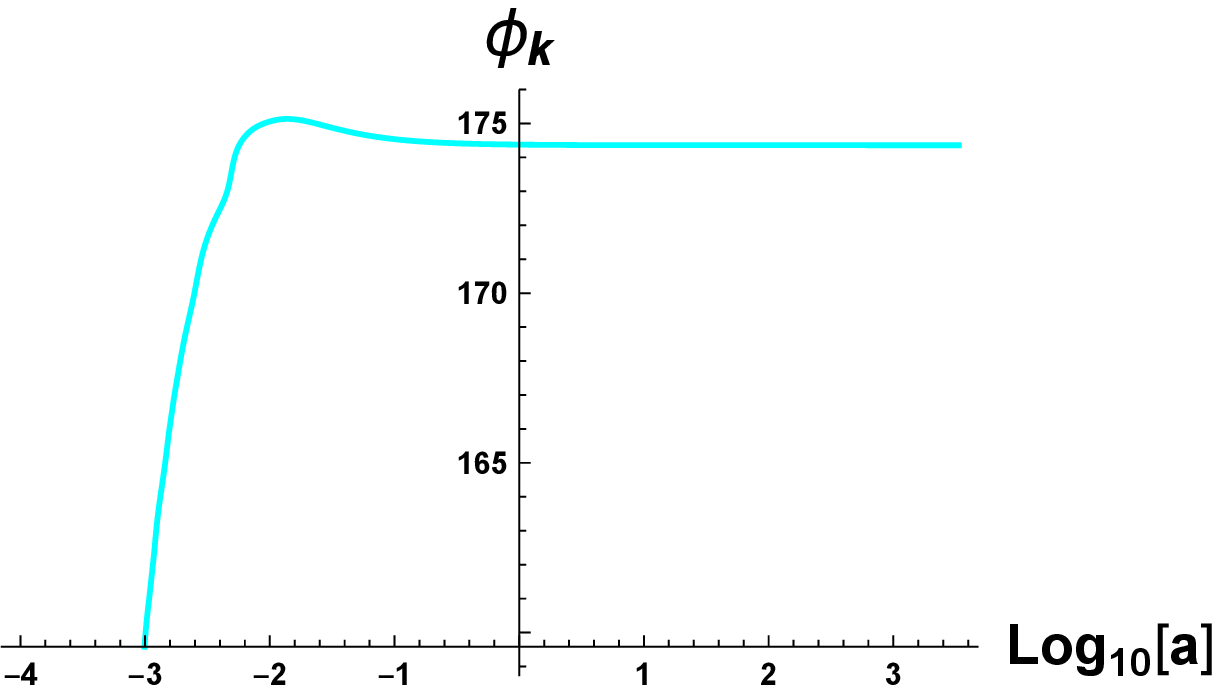}
		\caption{(color online). The squeezing parameters vs. the log of scale factor for the usual inflationary de Sitter phase. The left figure shows the squeezing parameter $r_k$ oscillates with small amplitude initially, and then it grows linearly with respect to $\log_{10} a$. Note that the sub-figure is the  amplification of the evolution of $r_k$ in early time before horizon exit. For the right figure, the squeezing angle $\phi_k$ grows intensively, then it keeps constant.}
		\label{squeera1}
	\end{center}
\end{figure}

\begin{figure}[ht]
	\begin{center}
		\includegraphics[scale=0.33]{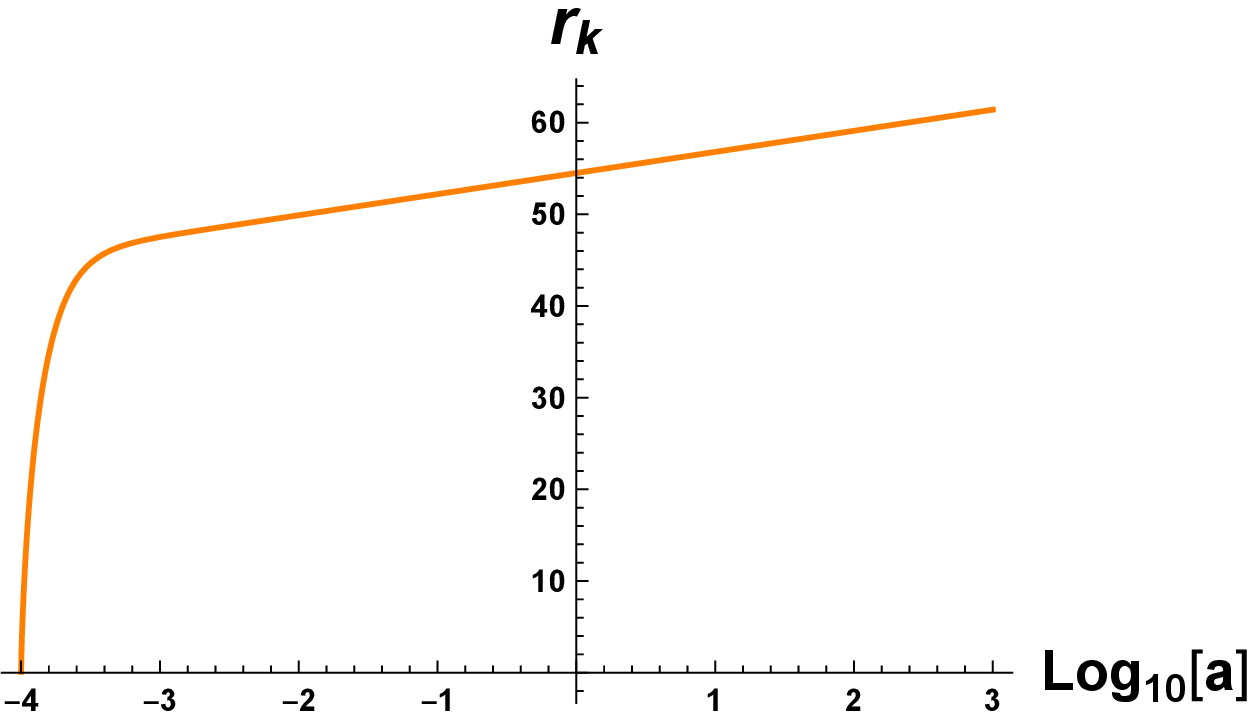}
		\includegraphics[scale=0.33]{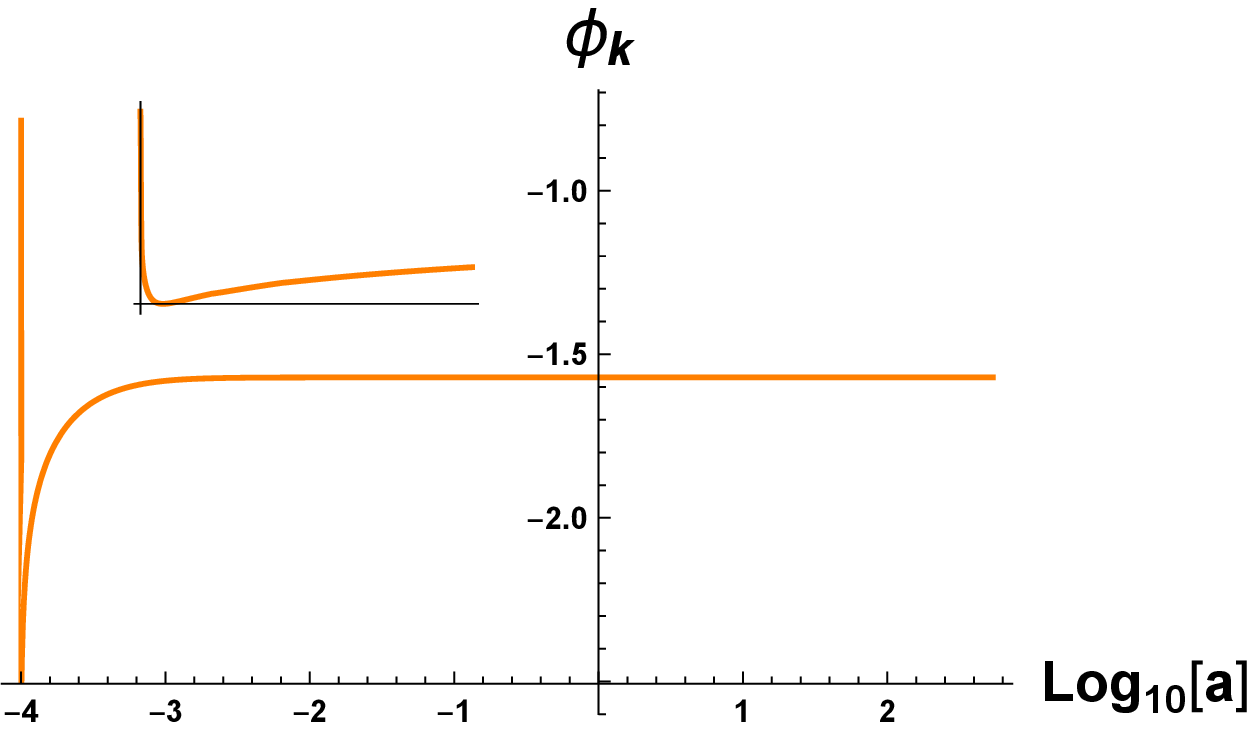}
		\caption{(color online). The squeezing parameters vs. the log of scale factor for the de Sitter expansion with the negative $c^2_S$. The left figure shows that the squeezing parameter $r_k$ increases dramatically, and then grows linearly with respect to $\log_{10} a$. For the right figure, the squeezing angle $\phi_k$ grows dramatically at first until it reaches the  minimal point, then it grows slowly. Note that the sub-figure is the amplification of $\phi_k$-$\log_{10}(a)$ in early time, which shows that the squeezing angle varies smoothly at the minimal point.}
		\label{squeera2}
	\end{center}
\end{figure}

\begin{figure}[ht]
	\begin{center}
		\includegraphics[scale=0.33]{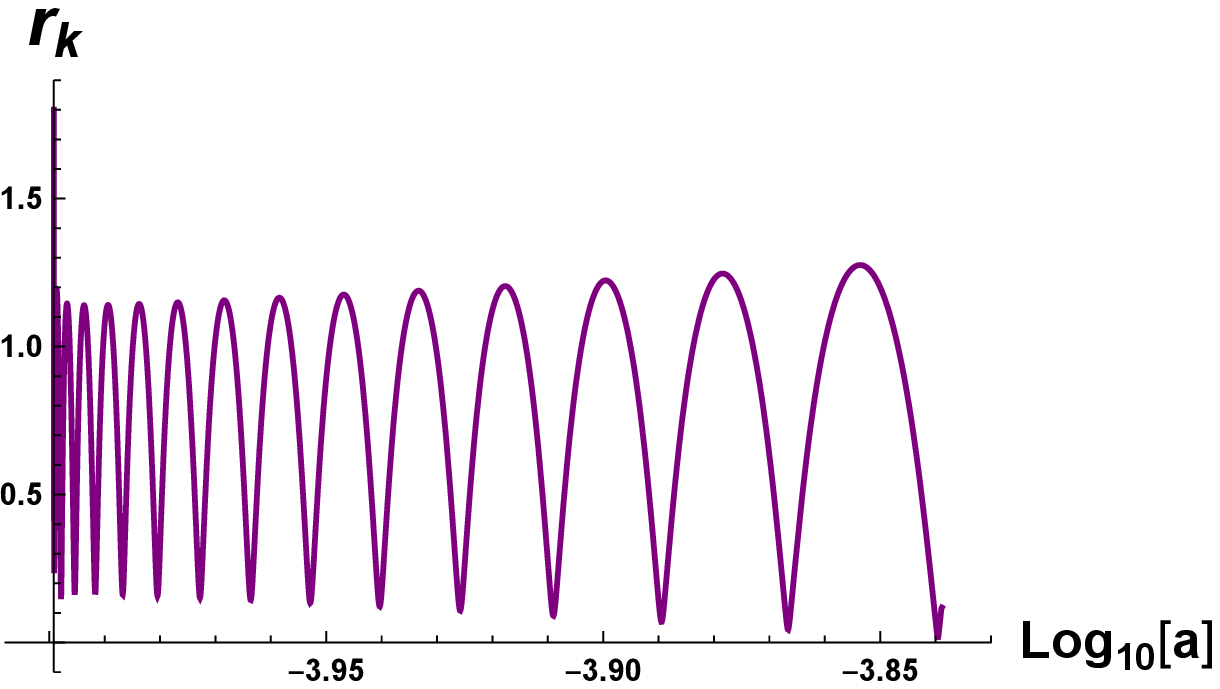}
		\includegraphics[scale=0.33]{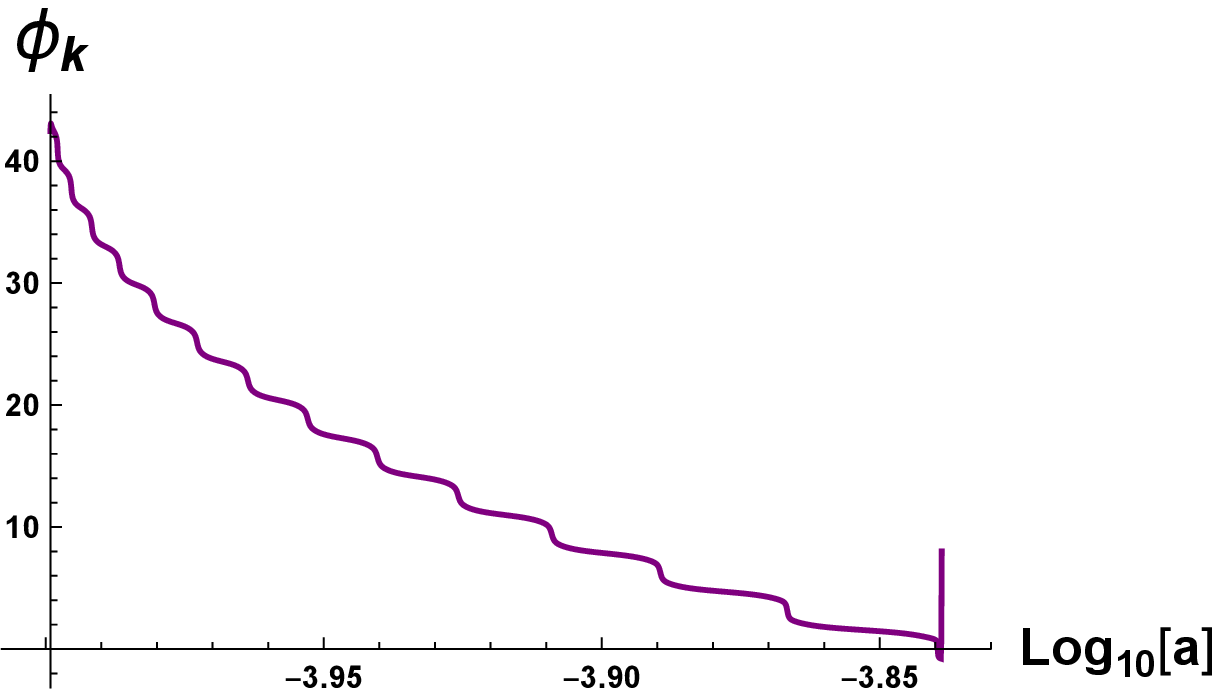}
		\caption{(color online). Squeezing parameters vs. the log of scale factor before bouncing. For the left figure, the magnitude of the squeezing parameter is almost constant while its frequency is increasing with respect to $\log_{10}$ as seen from the right to the left. For the right figure, the squeezing angle is gradually decreasing with small oscillation.}
		\label{squeera3BB}
	\end{center}
\end{figure}

\begin{figure}[ht]
	\begin{center}
		\includegraphics[scale=0.33]{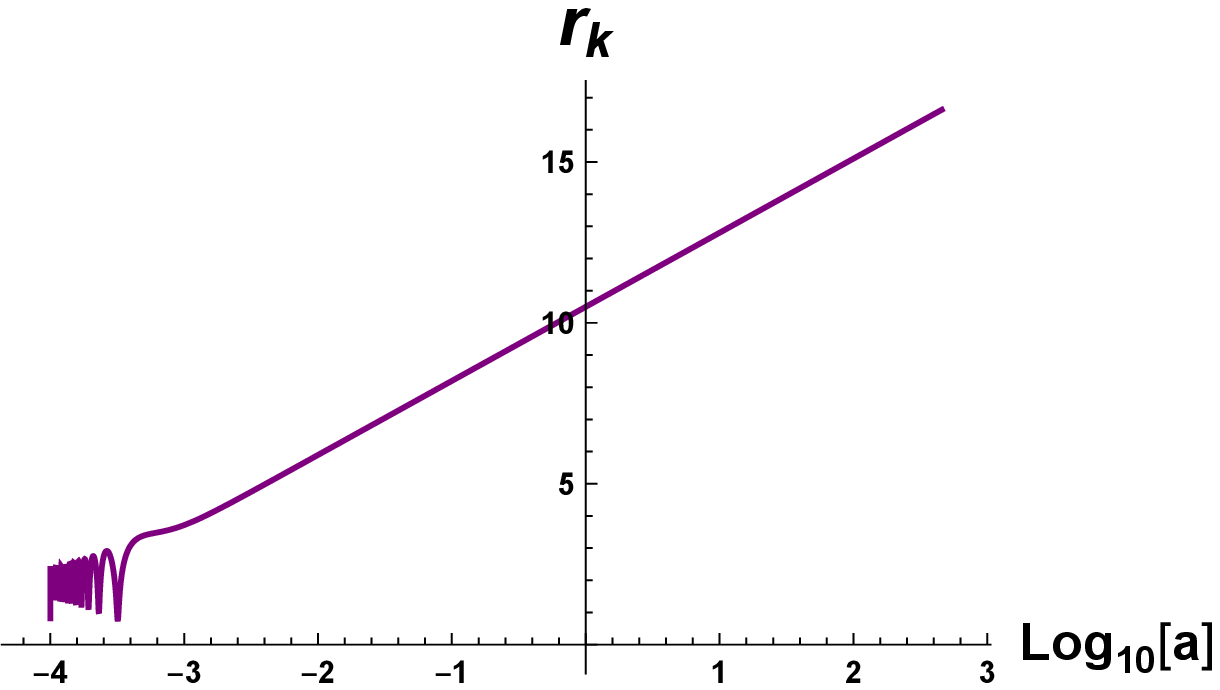}
		\includegraphics[scale=0.33]{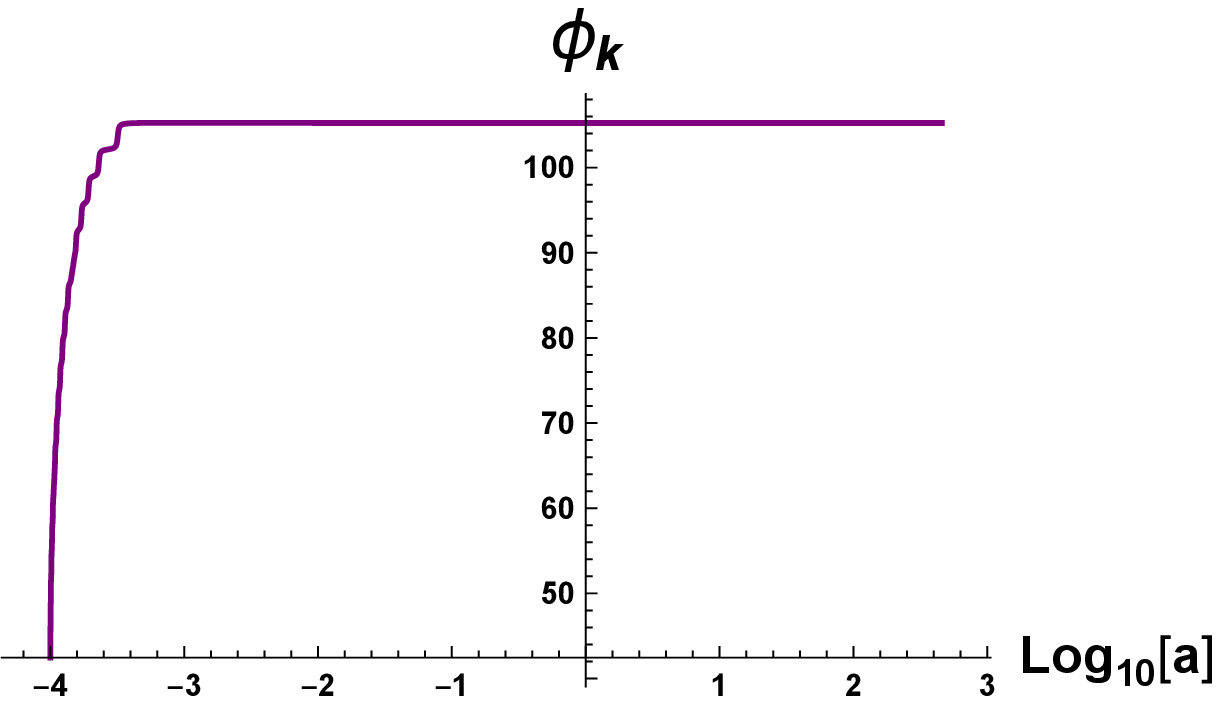}
		\caption{(color online). Squeezing parameters vs. the log of scale factor after bouncing. For the left figure, the magnitude of the squeezing parameter $r_k$ oscillates very fast at first, then its amplitude increases linearly with respect to $\log_{10}a$. For the right figure, the squeezing angle $\phi_k$ increases dramatically, then it tends to be constant.}
		\label{squeera3AB}
	\end{center}
\end{figure}

By using the operator ordering theorem given in \cite{TQO}, we could expand $\eqref{SqueeOpeVo}$ in the following ordered form
\begin{align}
\nonumber
&\hat{S}_{\vec{k}}(r_{k},\phi_{k})=\exp\big[-e^{2i\phi_{k}}\tanh r_{k}~\hat{c}_{-\vec{k}}^{\dagger}\hat{c}_{\vec{k}}^{\dagger}\big]\\
\nonumber
&\quad\quad\quad\quad~~~\cdot\exp\big[-\ln(\cosh r_{k})~\big(\hat{c}_{-\vec{k}}^{\dagger}\hat{c}_{-\vec{k}}+\hat{c}_{\vec{k}}\hat{c}_{\vec{k}}^{\dagger}\big)\big]\\
\label{SqueeOpeV1}
&\quad\quad\quad\quad~~~\cdot\exp\big[e^{-2i\phi_{k}}\tanh r_{k}~\hat{c}_{\vec{k}}\hat{c}_{-\vec{k}}\big]
\end{align}
After acting the squeeze operator $\eqref{SqueeOpeV1}$ on the two-mode vacuum state $\vert 0;0 \rangle_{\vec{k},-\vec{k}}$, a two-mode squeezed state will be got
\begin{align}
\label{squeeState}
&\vert\Psi\rangle_{sq}=\frac{1}{\cosh r_{k}}\sum_{n=0}^{\infty}(-1)^{n}e^{2in\phi_{k}}\tanh^{n}r_{k}\vert n;n\rangle_{\vec{k},-\vec{k}}
\end{align}
where the two-mode excited state $\vert n;n\rangle_{\vec{k},-\vec{k}}$ is
\begin{align}
&\vert n;n\rangle_{\vec{k},-\vec{k}}=\frac{1}{n!}\big(\hat{c}_{\vec{k}}^{\dagger}\big)^{n}\big(\hat{c}_{-\vec{k}}^{\dagger}\big)^{n}\vert0;0\rangle_{\vec{k},-\vec{k}}
\end{align}
After substituting $\eqref{HamWithCs},\eqref{squeeState}$ into the following $Schr\ddot{o}dinger$ equation
\begin{align}
&i\frac{d}{d\eta}\vert\Psi\rangle_{sq}=\hat{H} \vert\Psi\rangle_{sq}
\end{align}
we can give rise to the time evolution of the squeezing paremeters $r_k(\eta)~,\phi_k(\eta)$ 
\begin{align}
\label{evolurk}
&\hspace{-4mm}-\frac{dr_{k}}{d\eta}=\frac{k}{2}(c_{S}^{2}-1)\sin(2\phi_{k})+\frac{z^{\prime}}{z}\cos(2\phi_{k})\\
\nonumber
&\frac{d\phi_{k}}{d\eta}=\frac{k(c_{S}^{2}+1)}{2}-\frac{k}{2}(c_{S}^{2}-1)\cos2\phi_{k}\coth2r_{k}\\
\label{evoluphik}
&\quad\quad+\frac{z^{\prime}}{z}\sin2\phi_{k}\coth2r_{k}
\end{align}
Note that the variable $z$ is given by $\eqref{Zvariable}$, which is related to the scale factor $a$, the speed of sound $c^2_S$, the energy density $\mathcal{E}$ and the pressure $P$. For the sake of simplification in numerical calculations, the variable $\log_{10}(a)$ is used to take place of the conformal time $\eta$. After plugging the numerical solutions of $a(t),~c^2_S(t),~\mathcal{E}(t)+P(t)$ in Fig.\ref{EOMsXRHS}-Fig.\ref{EnerMoCSsqureSol} into the differential equations $\eqref{evolurk}$-$\eqref{evoluphik}$, we could show the evolution of squeezing parameters in Fig.\ref{squeera1}-Fig.\ref{squeera3AB}.
The result in Fig.\ref{squeera1} is analogous to the one given in \cite{Bhattacharyya:2020rpy}, the squeezing parameter $r_k$ stays small in sub-horizon limit and then grows rapidly in an exponential way after the mode exits the horizon.
The Fig.\ref{squeera2} corresponds to the numerical solutions of squeezing parameters in background of de Sitter expanding phase with negative $c^2_S$, in which the most remarkable character is that the squeezing parameter $r_k$ rises steeply without oscillatory behavior inside the horizon. This "fast-squeezed" behavior is originated from the negative $c^2_S$.
On the background of bounce phase, the evolutions of squeezing parameters in contracting and expanding stages are displayed in Fig.\ref{squeera3BB} and Fig.\ref{squeera3AB} respectively. This result is consist with \cite{Bhargava:2020fhl}, in which the circuit complexity of two well known bouncing cosmological solutions, i.e. $Cosine hyperbolic$ and $Exponential$ models of scale factors, have been studied. Similar to \cite{Bhargava:2020fhl}, in contracting stage, we observe that the squeezing parameter $r_k$ is vigorously oscillatory with decreasing amplitudes when approaching the bouncing point. After crossing the bouncing point and entering the expanding stage, the $r_k$ is oscillatory with slowly varying amplitudes at the beginning and then grows fast like the Fig.\ref{squeera1}. 

\section{Complexity for squeezed quantum states of cosmological perturbations \label{SqueeComplex}}

In this paper, we will evaluate the circuit complexity by using Nielsen's method \cite{NielsenComplexity1, NielsenComplexity2, NielsenComplexity3}. First, a reference state $\vert \psi ^R \rangle$ is given at $\tau=0$. And then, we suppose that a target state $\vert \psi ^T \rangle$ could be obtained at $\tau=1$ by acting a unitary operator on $\vert \psi ^R \rangle$, namely
\begin{align}
\label{ComWaveFunc}
&\vert \psi ^T \rangle_{\tau=1}=U(\tau=1) \vert \psi ^R \rangle_{\tau=0}
\end{align}
As usual, $\tau$ parametrizes a path in the Hilbert space. Generally, the unitary operator is constructed from a path-ordered exponential of a Hamiltonian operator
\begin{align}
&U(\tau)=\overleftarrow{\mathcal{P}} \exp \bigg( -i\int ^\tau _0 ds H(s) \bigg)
\end{align}
where the $\overleftarrow{\mathcal{P}}$ indicates right-to-left path ordering. The Hamiltonian operator $H(s)$ can be expanded in terms of a basis of Hermitian operators $M_I$, which are the generators for elementary gates
\begin{align}
&H(s)= Y(s)^I M_I
\end{align}
The coefficients $Y(s)^I$ are identified as the control functions that determine which gate should be switched on or switched off at a given parameter. Meanwhile, the $Y(s)^I$ satisfy the $Schr\ddot{o}dinger$ equation
\begin{align}
&\frac{dU(s)}{ds}=-iY(s)^I M_I U(s)
\end{align}
Then a $cost~functional$ is defined as follows
\begin{align}
\label{costfunctional}
&\mathcal{C}(U)=\int_0 ^1 \mathcal{F(U,\dot{U})}d\tau
\end{align}
The complexity is obtained by minimizing the functional $\eqref{costfunctional}$ and finding the shortest geodesic distance between the reference and target states. Here, we restrict our attentions on the $quadratic$ cost functional
\begin{align}
\label{quadfunctional}
&\mathcal{F}(U,Y)=\sqrt{\sum_I (Y^I)^2}
\end{align}
In this work, the target state is the two-mode squeezed vacuum state $\eqref{squeeState}$. After projecting $\vert\Psi\rangle$ into the position space, the following wavefunction is implied \cite{Martin:2019wta, Lvovsky:2014sxa}
\begin{align}
\nonumber
&\Psi_{sq}(q_{\vec{k}},q_{-\vec{k}})=\sum_{n=0}^{\infty}(-1)^{n}e^{2in\phi_{k}}\frac{\tanh^{n}r_{k}}{\cosh r_{k}}\langle q_{\vec{k}};q_{-\vec{k}}\vert n;n\rangle_{\vec{k},-\vec{k}}\\
\label{squeeposition}
&~=\frac{\exp[A(r_{k},\phi_{k})\cdot(q_{\vec{k}}^{2}+q_{-\vec{k}}^{2})-B(r_{k},\phi_{k})\cdot q_{\vec{k}}q_{-\vec{k}}]}{\cosh r_{k}\sqrt{\pi}\sqrt{1-e^{-4i\phi_{k}}\tanh^{2}r_{k}}}
\end{align}
in which the coefficients $A(r_k,\phi_k)$ and $B(r_k,\phi_k)$ are
\begin{align}
&A(r_{k},\phi_{k})=\frac{k}{2}\bigg(\frac{e^{-4i\phi_{k}}\tanh^{2}r_{k}+1}{e^{-4i\phi_{k}}\tanh^{2}r_{k}-1}\bigg)\\
&B(r_{k},\phi_{k})=2k\bigg(\frac{e^{-2i\phi_{k}}\tanh r_{k}}{e^{-4i\phi_{k}}\tanh^{2}r_{k}-1}\bigg)
\end{align}
By using a suitable rotation in vector space $(q_{\vec{k}},q_{-\vec{k}})$, the exponent in $\eqref{squeeposition}$ could be rewritten by a form of diagonal matrix
\begin{align}
\label{targerstate}
&\Psi_{sq}(q_{\vec{k}},q_{-\vec{k}})=\frac{\exp[-\frac{1}{2}\tilde{M}^{ab}q_{a}q_{b}]}{\cosh r_{k}\sqrt{\pi}\sqrt{1-e^{-4i\phi_{k}}\tanh^{2}r_{k}}}\\
\nonumber
&\tilde{M}=\left(\begin{array}{cc}
\Omega_{\vec{k}^{\prime}} & 0\\
0 & \Omega_{-\vec{k}^{\prime}}
\end{array}\right)=\left(\begin{array}{cc}
-2A+B & 0\\
0 & -2A-B
\end{array}\right)
\end{align}
Meanwhile, the reference state is the unsqueezed vacuum state,
\begin{align}
\nonumber
&\Psi_{00}(q_{\vec{k}},q_{-\vec{k}})=\langle q_{\vec{k}};q_{-\vec{k}}\vert0;0\rangle_{\vec{k},-\vec{k}}\\
\nonumber
&\quad\quad\quad\quad\quad~=\frac{\exp[-\frac{1}{2}(\omega_{\vec{k}}q_{\vec{k}}^{2}+\omega_{-\vec{k}}q_{-\vec{k}}^{2})]}{\pi^{1/2}}\\	
\label{referstate}	
&\quad\quad\quad\quad\quad~=\frac{\exp[-\frac{1}{2}\tilde{m}^{ab}q_aq_b]}{\pi^{1/2}}\\
\nonumber
&\tilde{m}=\left(\begin{array}{cc}
\omega_{\vec{k}} & 0\\
0 & \omega_{-\vec{k}}
\end{array}\right)
\end{align}
According to the definition $\eqref{ComWaveFunc}$, one can associate the target state $\eqref{targerstate}$ with the reference state $\eqref{referstate}$ through a unitary transformation 
\begin{align}
\label{Utrans1}
&\Psi_\tau(q_{\vec{k}},q_{-\vec{k}})=\tilde{U}(\tau)\Psi_{00}(q_{\vec{k}},q_{-\vec{k}}) \tilde{U}^\dagger (\tau)\\
\label{boun1}
&\Psi_{\tau=0}(q_{\vec{k}},q_{-\vec{k}})=\Psi_{00}(q_{\vec{k}},q_{-\vec{k}})\\
\label{boun2}
&\Psi_{\tau=1}(q_{\vec{k}},q_{-\vec{k}})=\Psi_{sq}(q_{\vec{k}},q_{-\vec{k}})
\end{align}
where $\tilde{U}(\tau)$ is a $GL(2,C)$ unitary matrix which give the shortest geodesic distance between the target state and the reference state in operator space. In general, the operators of $GL(2,C)$ can be expressed as
\begin{align}
&\tilde{U}(\tau)=\exp[\sum_{I=1}^{4}Y^I (\tau) M_I]
\end{align}
where the $\{M_I\}$ represent the 4 generators of $GL(2,C)$, namely
\begin{align}
\nonumber
&M_1=\left(\begin{array}{cc}
1 & 0\\
0 & 0
\end{array}\right)~,~M_2=\left(\begin{array}{cc}
0 & 0\\
0 & 1
\end{array}\right)
\\
\nonumber
&M_3=\left(\begin{array}{cc}
0 & 1\\
0 & 0
\end{array}\right)~,~M_4=\left(\begin{array}{cc}
0 & 0\\
1 & 0
\end{array}\right)
\end{align}
Therefore the geometry in this operator space is described by the metric \cite{Chapman:2017rqy},
\begin{align}
&ds^2=G_{IJ}dY^I d{(Y^J)^{\star}}
\end{align} 
In language of geometry, together with definitions $\eqref{costfunctional}$ and $\eqref{quadfunctional}$, the complexity is described by the following line length
\begin{align}
\label{CompleLineLeng}
&C(\tilde{U})=\int^1_0 d\tau \sqrt{G_{IJ}\dot{Y}^I(\tau) ({\dot{Y}}^J(\tau))^\star}
\end{align}
in which the dot denotes the derivative with respect to $t$. 
Due to the fact that the manifold of general linear group $GL(N,C)$ could be an Euclidean geometry, it is simple to set $G_{ij}=\delta_{ij}$. It follows that the  shortest geodesic between target state and reference state is a straight line, namely
\begin{align}
\label{strailineYtau}
&Y^I (\tau)=Y^I (\tau=1)\cdot \tau+ Y^I (\tau=0)
\end{align}
Note that both $\eqref{targerstate}$ and $\eqref{referstate}$ are diagonal, the off-diagonal generators will increase the distance between two states in operator space; thus the components $Y^3, Y^4$ is set to zero \cite{Jefferson:2017sdb}. With the help of boundary conditions $\eqref{boun1}$ and $\eqref{boun2}$, one gets
\begin{align}
\label{Ytau0}
&\quad~\text{Im}(Y^{1,2})\big\vert_{\tau=0}=\text{Re}(Y^I)\big\vert_{\tau=0}=0\\
\label{Ytau1}
&\begin{cases}
\begin{array}{c}
\hspace{-10mm}\text{Im}(Y^{1,2})\big\vert_{\tau=1}=\frac{1}{2}\ln\frac{\vert\Omega_{\vec{k},\vec{-k}}\vert}{\omega_{\vec{k},\vec{-k}}}\\
\text{Re}(Y^{1,2})\big\vert_{\tau=1}=\frac{1}{2}\arctan\frac{\text{Im}(\Omega_{\vec{k},\vec{-k}})}{\text{Re}(\Omega_{\vec{k},\vec{-k}})}
\end{array}\end{cases}
\end{align}
Finally, plugging $\eqref{strailineYtau}$-$\eqref{Ytau1}$ into $\eqref{CompleLineLeng}$, the complexity is calculated as
\begin{widetext}
\begin{align}
&\mathcal{C}(k)=\frac{1}{2}\sqrt{\big(\ln\frac{\vert\Omega_{\vec{k}}\vert}{\omega_{\vec{k}}}\big)^{2}+\big(\ln\frac{\vert\Omega_{-\vec{k}}\vert}{\omega_{-\vec{k}}}\big)^{2}+\big(\arctan\frac{\text{Im}(\Omega_{\vec{k}})}{\text{Re}(\Omega_{\vec{k}})}\big)^{2}+\big(\arctan\frac{\text{Im}(\Omega_{\vec{-k}})}{\text{Re}(\Omega_{\vec{-k}})}\big)^{2}}
\end{align}
\end{widetext}
where $\omega_{\vec{k}}=\omega_{-\vec{k}}=\vert \vec{k} \vert=k$.

\begin{figure}[ht]
	\begin{center}
		\includegraphics[scale=0.33]{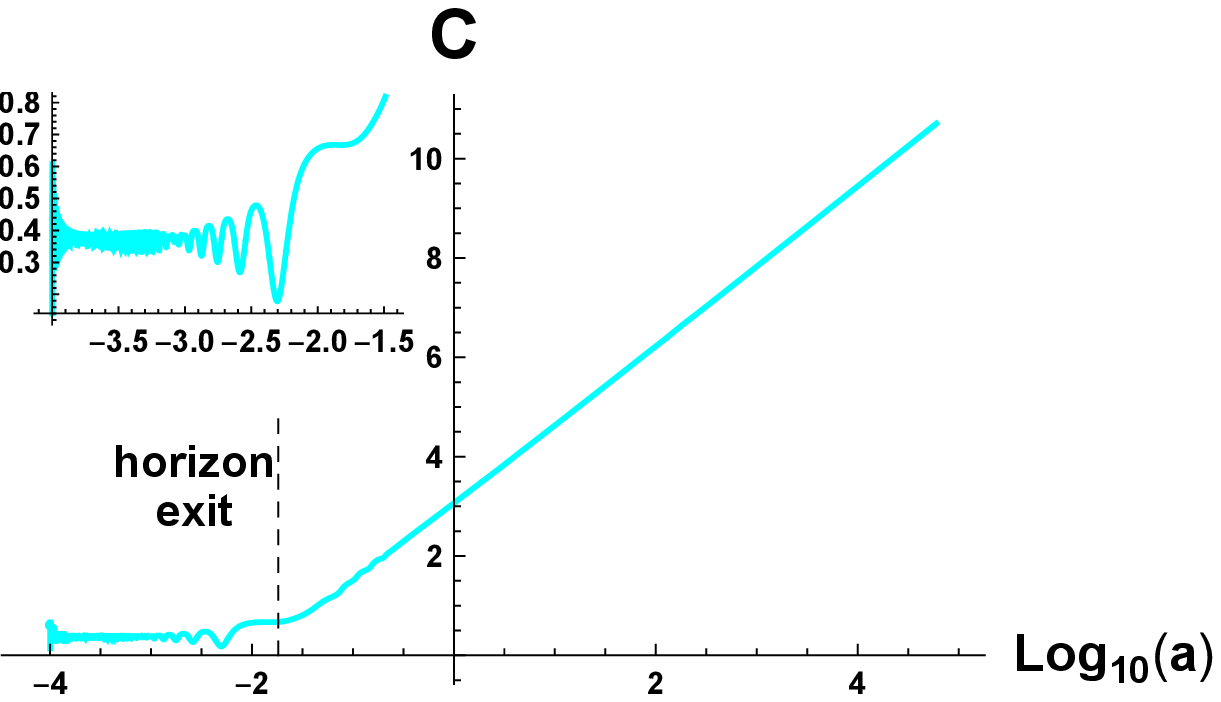}
		\includegraphics[scale=0.33]{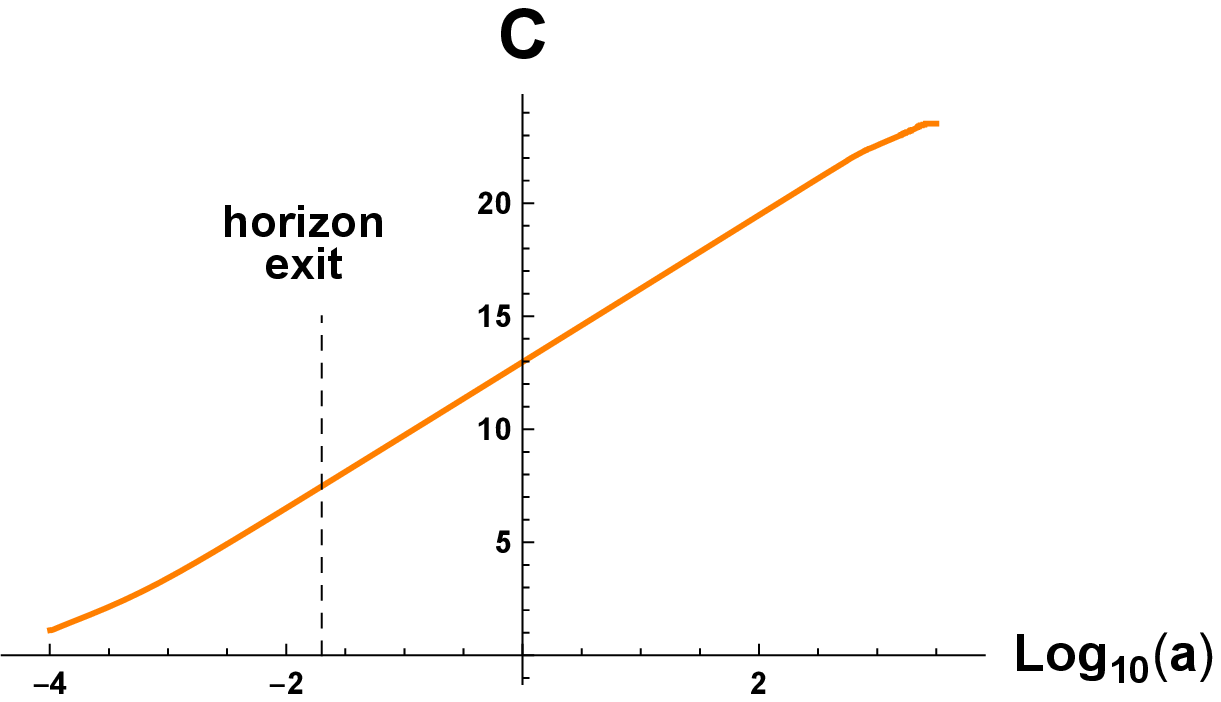}
		\caption{(color online). Complexity vs. the log of scale factor. For the left figure in the de-Siiter inflation, the complexity oscillates at first, and its amplitude keeps constant. Then, the complexity grow linearly with respect to $\log_{10} a$. For the right figure in the inflation with the squire of sound speed $c^2_s<0$, the complexity grows linearly.}
		\label{squeeCom12}
	\end{center}
\end{figure}
\begin{figure}[ht]
	\begin{center}
		\includegraphics[scale=0.33]{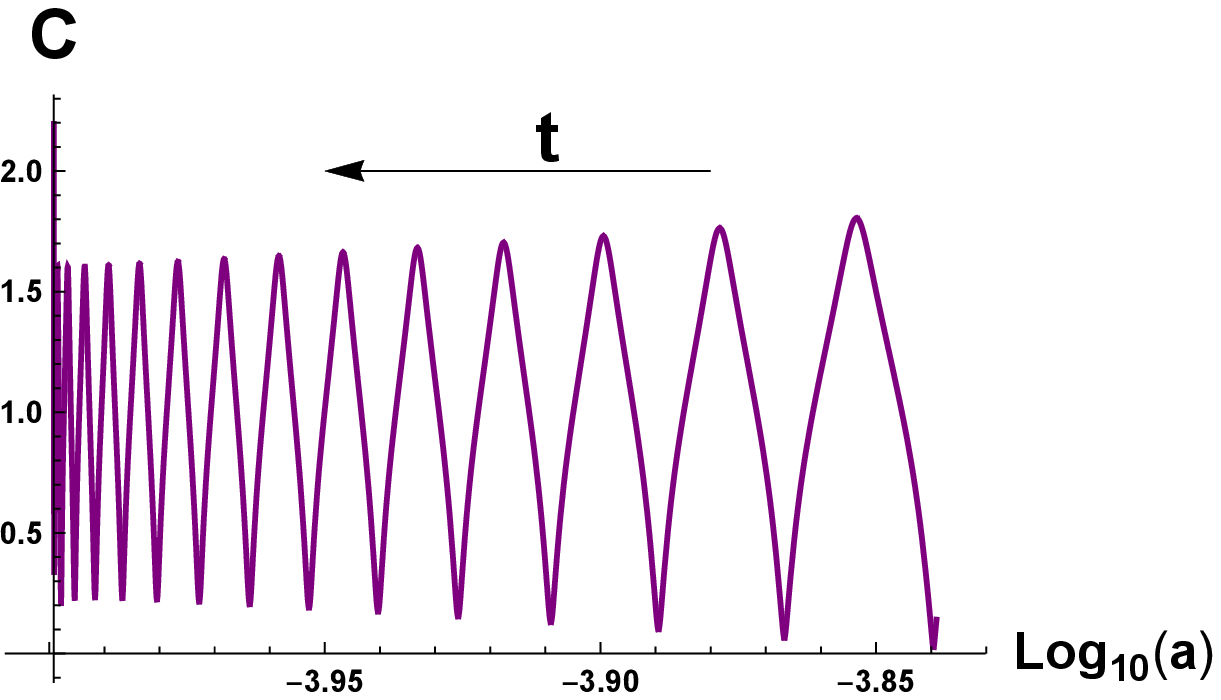}
		\includegraphics[scale=0.33]{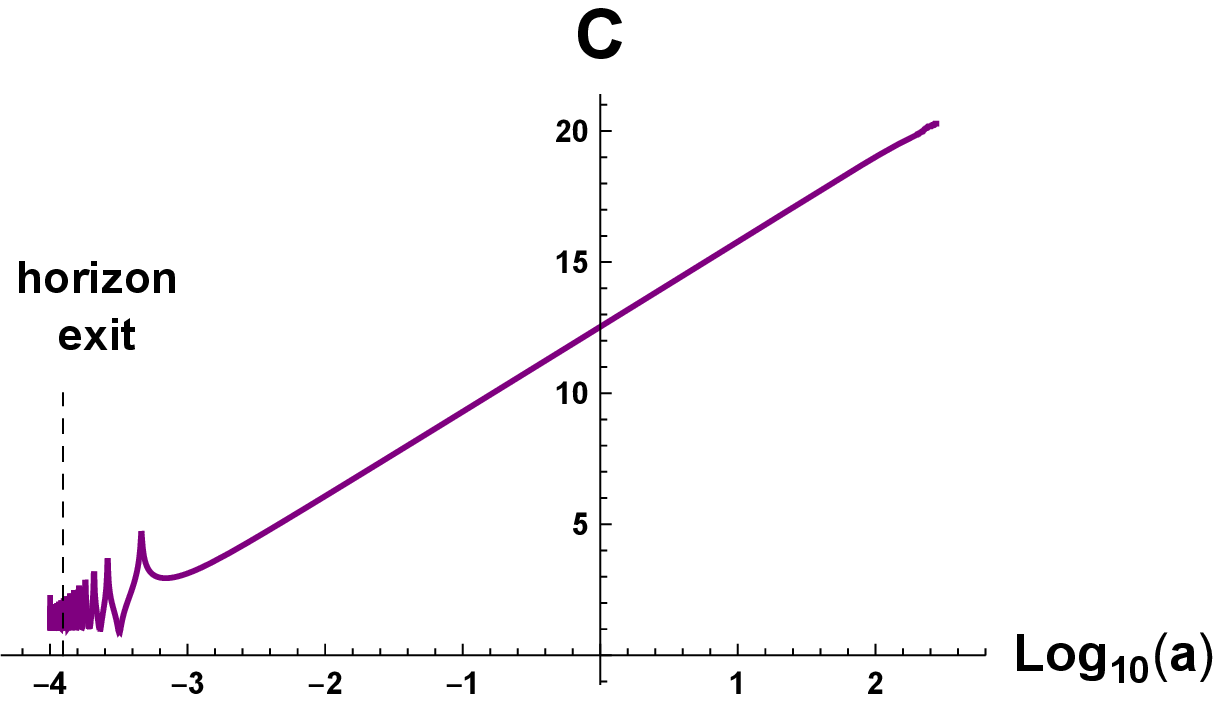}
		\caption{(color online). Complexity vs. the log of scale factor. For the left figure, before bouncing, the complexity oscillates, and its frequency keeps growing. For the right figure, after bouncing, the complexity oscillates intensively at first, then its complexity grows linearly with respect to $\log_{10} a$}
		\label{squeeCom3}
	\end{center}
\end{figure}
The complexity is explicitly obtained in Fig.\ref{squeeCom12}-Fig.\ref{squeeCom3} via the numerical solutions of $r_k$ and $\phi_k$. As indicated by \cite{Ali:2019zcj}, some basic informations about a quantum chaotic system could also be captured by the circuit complexity. In particular, by studying an inverted harmonic oscillator model, they find that the time scale when the complexity starts to grow could be identified as the scrambling time, while the Lyapunov exponent approximately equals to the slope of the linear growth part. Basing on the analysis of \cite{Ali:2019zcj}, after calculating the cosmological complexity on background of inflationary de Sitter expansion, \cite{Bhattacharyya:2020rpy} finds that the scrambling time scale is the time of horizon exit while the $\frac{dC}{dt}$ is approximately equal to the Hubble constant (to be precise, $\frac{dC}{dt}<H$). Their results are consistent with the left panel of Fig.\ref{squeeCom12}, which describes the evolution of the cosmological complexity for K-essence in the usual inflationary de Sitter phase. However, in case of the de Sitter expansion with negative $c^2_S$, the scrambling time appears in the subhorizon limit while the changing rate of the complexity in linear growth part satisfies $\frac{dC}{dt}<H$, as displayed by the right panel of Fig.\ref{squeeCom12}. Finally, for the bouncing cosmology phase, one should note in Fig.\ref{squeeCom3} that the scrambling time occurs after the time of horizon exit, and the $\frac{dC}{dt}<H$ in linear growth region is smaller than the value of Hubble constant.

\section{Conclusions and Discussion \label{ConDiss}}

In this paper, we have computated the cosmological complexity in a type of K-essence cosmology model with constant potential \cite{Jorge:2007zz}. After solving the dynamical system of this K-essence model, as shown in Fig.\ref{EOMsXRHS}-Fig.\ref{EnerMoCSsqureSol}, we observe three kinds of cosmological phases in same physical parameters with different initial conditions , namely the usual inflationary de Sitter phase (cyan curves), the de Sitter expansion with negative $c^2_S$ (orange curves), the bouncing cosmology phase (purple curves). Besides, basing on the work \cite{Garriga:1999vw}, the perturbative action describing the perturbations of curvature scalar is given by $\eqref{PerLagranKessen}$, in which the varied $c^2_S$ is included due to the higher order corrections to the kinetic term of scalar field $\varphi$. By utilizing the method of second quantization to the perturbative action, together with the definition of squeezed quantum state which is characterized by squeezing parameter $r_k$ and squeezing angle $\phi_k$, we obtain the differential equations governing the evolution of $r_k$ and $\phi_k$ in $\eqref{evolurk}$-$\eqref{evoluphik}$. The numerical solutions of these differential equations in backgrounds of three kinds of cosmological phases are displayed in Fig.\ref{squeera1}-Fig.\ref{squeera3AB} respectively. Subsequently, the physical meaning of quantum circuit complexity and a type of computational method called wave-function approach are reviewed briefly. According to this approach, the complexity between unsqueezed vacuum state and squeezed quantum states are calculated under the framework of cosmological perturbations. The evolution of cosmological complexity in different cosmological phases are shown from Fig.\ref{squeeCom12} to Fig.\ref{squeeCom3}. 

For the evolution of cosmological complexity in the usual inflationary de Sitter phase, we observe the consistent results compared to the work \cite{Bhattacharyya:2020rpy}, namely the scrambling time scale is just the time of horizon exit while the $\frac{dC}{dt}$ in linear growth portion is in same magnitude as the Hubble constant (to be precise, $dC/dt<H$). However, for the de Sitter expansion with negative $c^2_S$, the scrambling time occurs far earlier than the time of horizon exit. And the changing rate of the complexity in linear growth part is bigger than the value of Hubble constant. Besides, it is easy to observe that the complexity grows rapidly without any oscillatory behavior in early state inside the horizon. Actually, this fast-growing phenomenon is arisen from the negative $c^2_S$ which leads to the "fast-squeezed" behavior of the $r_k$ in subhorizon limit as shown by Fig.\ref{squeera2}. Finally, in regard to the bouncing cosmology phase, the scrambling time is lag behind the time of horizon exit while the changing rate of the complexity in linear growth part satisfies $dC/dt>H$. As a character in bouncing cosmology phase, it should be noticed that the oscillatory behavior of the cosmological complexity would last after the horizon exit.

As discussions, we suggest the following extended topics. In this paper, we estimate the scarambling time and Lyapunov exponent just basing on the analysis of \cite{Ali:2019zcj}. Actually, one can more precisely obtain these physical quantities by applying the out-of-time-order correlator (OTOC) method \cite{Haque:2020pmp} to this K-essence cosmology model. Besides, one could also apply the cosmological complexity to investige the multi-field inflation model. Especially, it is interesting to explore the possible associations between the geometrical instability arising in multi-field inflation and the Lyapunov exponent estimated from the cosmological complexity or OTOC.

\section{Acknowledgements}
AC and DF are supported by NSFC grant no.11875082. Meanwhile, AC is also supported by the University of Barcelona (UB) / China Scholarship Council (CSC) joint scholarship. XF is supported by the Doctor Start-up Foundation of Guangxi University of Science and Technology with Grant No.19Z21. LH is funded by Hunan Natural Provincial Science Foundation NO. 2020JJ5452 and Hunan Provincial Department of Education, NO. 19B46.

\end{document}